\PassOptionsToPackage{pdfpagelabels=false}{hyperref}

\documentclass[fleqn,usenatbib,useAMS]{mnras}

\usepackage{graphicx}
\usepackage{amsmath}
\usepackage[dvipsnames]{xcolor}  
\usepackage{amssymb}
\usepackage{color}
\usepackage[figure,figure*,table,table*]{hypcap}
\usepackage[normalem]{ulem}
\usepackage{soul}

\input{macros.sty}

\title[Satellite Kinematics in SDSS]
{Updated Results on the Galaxy-Halo Connection from Satellite Kinematics in SDSS}
\author[J.~U.~Lange et al.]
{Johannes~U.~Lange$^1$\thanks{email: johannesulf.lange@yale.edu}, Frank~C.~van~den~Bosch$^1$, Andrew~R.~Zentner$^2$\newauthor Kuan~Wang$^2$ and Antonio~S.~Villarreal$^3$\\
	$^1$Department of Astronomy, Yale University, P.O. Box 208101, New Haven, CT 06511, USA\\
	$^2$Department of Physics and Astronomy \& Pittsburgh Particle Physics, Astrophysics, and Cosmology Center (PITT PACC),\\University of Pittsburgh, Pittsburgh, PA 15260, USA\\
	$^3$Argonne National Laboratory, Argonne, IL 60439, USA}

\pubyear{2018}

\begin{document}
	
	\date{Accepted xxx. Received xxx}
	
	\label{firstpage}
	\pagerange{\pageref{firstpage}--\pageref{lastpage}}
	
	\maketitle
	
	\begin{abstract}
		We present new results on the relationship between central galaxies and dark matter haloes inferred from observations of satellite kinematics in the Sloan Digital Sky Survey (SDSS) DR7. We employ an updated analysis framework that includes detailed mock catalogues to model observational effects in SDSS. Our results constrain the colour-dependent conditional luminosity function (CLF) of dark matter haloes, as well as the radial profile of satellite galaxies. Confirming previous results, we find that red central galaxies live in more massive haloes than blue galaxies at fixed luminosity. Additionally, our results suggest that satellite galaxies have a radial profile less centrally concentrated than dark matter but not as cored as resolved subhaloes in dark matter-only simulations. Compared to previous works using satellite kinematics by More et al., we find much more competitive constraints on the galaxy-halo connection, on par with those derived from a combination of galaxy clustering and galaxy-galaxy lensing. We compare our results on the galaxy-halo connection to other studies using galaxy clustering and group catalogues, showing very good agreement between these different techniques. We discuss future applications of satellite kinematics in the context of constraining cosmology and the relationship between galaxies and dark matter haloes.
	\end{abstract}
	
	
	\section{Introduction}
	
	In the standard cosmological paradigm, dark matter haloes provide the potential wells for baryonic material to condense and, ultimately, form galaxies. Thus, the relationship between galaxies and dark matter haloes, commonly called the galaxy-halo connection \citep[see, e.g.][and references therein]{Wechsler_18}, is central to our understanding of galaxy formation. Of particular interest is the relationship between galaxy properties and dark matter halo mass.
	
	There exist numerous techniques to observationally infer the average dark matter halo mass of galaxies. Among the observables used to constrain this relationship are galaxy clustering \citep[e.g.,][]{Yang_03, Zehavi_05, Zehavi_11, vdBosch_07, Zheng_07, Hearin_13c, Guo_15a, Guo_15b, Guo_16, Zentner_16, Xu_18}, galaxy-galaxy lensing \citep[e.g.,][]{Leauthaud_12, Zu_15, Zu_16, Mandelbaum_16, Leauthaud_17, Sonnenfeld_18a}, and galaxy group catalogues \citep[][]{Weinmann_06, Yang_07, Yang_08, Yang_09, Kauffmann_13, Hoshino_15, Sinha_18}. While those observations provide meaningful constraints on the galaxy-halo connection, they also suffer from a variety of systematic issues, such as halo assembly bias \citep{Gao_05, Wechsler_06, Zentner_14, Zentner_16, Villarreal_17}, miscentering \citep{Johnston_07, Skibba_11, Hikage_13, Hoshino_15, Lange_18a}, spectroscopic incompleteness \citep{Blanton_03a, Zehavi_11, Guo_12b}, and group finder errors \citep{Campbell_15, Zu_17, Calderon_18} among others. Thus, to maximise the information content of existing and future galaxy surveys, it is important to pursue additional ways to constrain the galaxy-halo connection.
	
	In \citet[][hereafter Paper I]{Lange_18b}, we presented an updated framework to extract the galaxy-halo connection from observations of satellite kinematics \citep[e.g.,][]{Zaritsky_93, Brainerd_03, Prada_03, vdBosch_04, More_09a, More_09b, More_11, Wojtak_13}. The idea of such an analysis is to use satellite galaxies as dynamical tracers of the host dark matter halo potential. By stacking the signal of a large number of isolated galaxies, such an analysis can determine average halo masses even for central galaxies that host on average only a few satellites. In Paper I, we showed how to correct for important observational biases like fibre collisions and demonstrated the robustness of those constraints by applying our technique to mock catalogues with various degrees of complexity. Furthermore, we demonstrated that constraints derived with this new technique are competitive even when compared to combined studies of clustering and lensing \citep[e.g.,][]{Cacciato_13, Zu_15, Zu_16}.
	
	The main goal of this analysis is to apply our analysis framework to the \textit{Sloan Digital Sky Survey} \citep[][hereafter SDSS]{York_00}. Specifically, we constrain the colour-dependent conditional luminosity function \citep{Yang_03} of dark matter haloes. We compare our results on the galaxy occupation to those results derived from clustering and group catalogues, finding them to be in very good agreement. Compared to the previous analysis of satellite kinematics in the SDSS by \cite{More_09b, More_11}, our analysis accounts for all relevant observational effects and biases, including fibre collisions. As we demonstrate, this removes the tension in the inferred galaxy-halo connection evident from a comparison of previous satellite kinematics studies with results inferred from alternative methods \citep{More_11, Leauthaud_12, Mandelbaum_16, Wechsler_18}. In addition, our results also constrain the radial profiles of satellite galaxies in dark matter haloes. We show that satellites are spatially anti-biased with respect to dark matter but are not as centrally cored as subhaloes in dark matter-only simulations.
	
	This paper is organised as follows. In \S\ref{sec:data}, we describe the observational data used to constrain the galaxy-halo connection. We summarise our modelling ingredients in \S\ref{sec:model}. Our detailed analysis procedure, including the use of mock catalogues to calibrate our analytic model and calculate covariance matrices, is described in \S\ref{sec:analysis}. In \S\ref{sec:results} we present our results on the galaxy-halo connection and the radial profile of satellites, which we compare to results from independent studies in \S\ref{sec:discussion}. Finally, we summarise our results in \S\ref{sec:conclusion}.
	
	Throughout this work, we assume a $\Lambda$CDM cosmology with $\Omega_m = 0.3071$, $\Omega_b = 0.0483$, $n_s = 0.9611$, $\sigma_8 = 0.8288$ and $h = H_0 / 100 \mathrm{km/s/Mpc} = 0.6777$, the best-fit results from the cosmic microwave background analysis of \cite{Planck_14}. All magnitudes are given in the AB magnitude system. Additionally, throughout this paper we use $r$ to denote 3D radii, and $R$ for projected 2D radii.
	
	\section{Observational Data}
	\label{sec:data}
	
	Observational constraints in this work come from the New York University Value-Added Galaxy Catalog \citep[VAGC;][]{Blanton_05}. This catalogue is derived from the Seventh Data Release of the SDSS \citep[SDSS DR7;][]{Abazajian_09}. Specifically, we use the \texttt{bright0} sample\footnote{\url{http://sdss.physics.nyu.edu/lss/dr72/bright/0/}} of the VAGC. This includes roughly $\sim 570,000$ galaxies with a limiting Petrosian magnitude of $m_r < 17.6$. $k$-corrections and evolution corrections to redshift $z = 0.1$ have been applied to all galaxies in the sample. Our analysis framework is optimised for volume-limited samples, as described in Paper I. We thus limit our analysis to galaxies with $L_r^{0.1} > 10^{9.5} \Lsunh$ and $0.02 < z < 0.067$. We apply the same $g - r$ colour cut to divide galaxies into red and blue as used by \cite{Zehavi_11}, written as
	\begin{equation}
		(g - r)_{\rm cut}^{0.1} = 0.21 - 0.03 (M_r^{0.1} - 5 \log h).
	\end{equation}
	Note that this cut is slightly different than the cut \cite{More_11} applied. Due to mechanical limitations, objects separated by less than $55''$ cannot both be assigned spectroscopic fibres on a single spectroscopic plate. Due to these fibre collisions around $\sim 6\%$ of the targets lack a spectroscopic redshift. In this case, we use the nearest-neighbour redshift assignment scheme \citep{Zehavi_05}. Note, though, that we only use these fibre-collided galaxies during the selection of primaries; they are not used when analysing the kinematics.
	
	\subsection{Sample Selection}
	
	As described in Paper I, we use a cylindrical isolation criterion to identify central candidates. A galaxy is considered isolated, a primary, if it does not have another brighter galaxy within a cylindrical volume defined by depth $(\Delta v)_\rmh$ and radius $R_\rmh$. We make this assessment using a list of all galaxies rank-ordered by luminosity, starting with the brightest. Any other fainter galaxy inside this cylindrical volume is removed from the list of potential primaries. The cylinder dimensions, as described below, are a function of the luminosity and colour of the galaxy in question. We remove galaxies close to the angular survey edge. This is characterised by putting a ring around each galaxy corresponding to the angular size of $R_\rmh$ at that galaxy's redshift and demanding that not more than $20\%$ of it lies outside the survey area. Finally, primaries are required to lie in a survey region with at least $80\%$ spectroscopic completeness. Afterwards, satellite candidates, so called secondaries, lying inside a cylinder defined by $(\Delta v)_\rms$ and $R_\rms$ are associated to each primary. Secondaries are allowed to lie outside the nominal redshift range, $0.02 < z < 0.067$. Contrary to \cite{vdBosch_04} and \cite{More_09b, More_09a, More_11}, primaries are not removed if they have no secondary. Doing so allows us to fold information about the fraction of centrals hosting satellites into the analysis. To limit the effect of fibre collisions, we remove all secondaries that are within $60 \kpch$, roughly corresponding to $55''$ at $z = 0.067$, of the primary (see Paper~I for details).
	
	We follow \cite{vdBosch_04} and make the cylinder sizes dependent on the galaxy properties. This typically increases completeness and purity of the resulting sample of central and satellite candidates \citep{vdBosch_04}. Specifically, we choose $(\Delta v)_\rmh = 1000 \sigma_{200} (L, C) \kms$, $R_\rmh = 0.5 \sigma_{200} (L, C) h^{-1} \Mpc$, $(\Delta v)_\rms = 4000 \kms$ and $R_\rms = 0.15 \sigma_{200} (L, C) h^{-1} \Mpc$. Here, $\sigma_{200} = \sigma / 200 \kms$ is an estimate for the velocity dispersion of satellites as a function of primary luminosity $L$ and colour $C$. In order to obtain similar completeness levels for red and blue centrals, we set $(\Delta v)_\rmh$ and $R_\rmh$ of blue centrals to those of red centrals of the same luminosity. Because the estimate for $\sigma_{200}$ is a priori unknown, the cylinder sizes must be extracted from the data itself in an iterative fashion. We start with an estimate for $\sigma_{200} (L, C)$ and extract primaries and secondaries. An estimate for $\sigma_{200} (L, C)$ is extracted from the data by fitting the distribution of secondaries in the $\Delta v_z$-$L_{\rm pri}$ plane. Here, $\Delta v_z$ is the line-of-sight velocity difference of the secondary and its primary and $L_{\rm pri}$ the luminosity of the primary. The model is
	\begin{equation}
		P(\Delta v_z, L_{\rm pri}) = \frac{f_{\rm int}}{2 (\Delta v)_\rms} + \frac{1 - f_{\rm int}}{\sqrt{2 \pi \sigma^2} \mathrm{erf}\left[\frac{(\Delta v)_\rms}{\sqrt{2} \sigma}\right]} \exp \left[ - \frac{\Delta v_z^2}{2 \sigma^2} \right],
	\end{equation}
	where both the interloper fraction $f_{\rm int}$ and the velocity dispersion $\sigma$ are a function of the primary luminosity. Specifically, $f_{\rm int}$ is assumed to be a linear function of $\log L_{\rm pri}$ and
	\begin{equation}
		\log \sigma / \kms = a + b (\log L_{\rm pri} - 10) + c (\log L_{\rm pri} - 10)^2,
	\end{equation}
	where $a$, $b$ and $c$ are constants. The best fit is found by maximising the likelihood,
	\begin{equation}
		\log \mathcal{L} \sim \sum\limits_{\mathrm{secondaries}} \log P(\Delta v_{z, i}, L_{\mathrm{pri}, i})^{w_{\mathrm{sw}, i}},
		\label{eq:likelihood_mem}
	\end{equation}
	where $w_{\mathrm{sw}, i}$ is a weight for each primary-secondary pair, as discussed in the next section. We then update our estimate for $\sigma_{200} (L, C)$ and extract a new set of primaries and secondaries. Using the best-fit values of \cite{More_11} as a a starting point, we find our algorithm to arrive at $a = 2.210$, $b = 0.478$, $c = 0.275$ and $a = 2.142$, $b = 0.402$, $c = -0.170$ for red and blue galaxies, respectively.
	\begin{figure}
		\centering
		\includegraphics[width=\columnwidth]{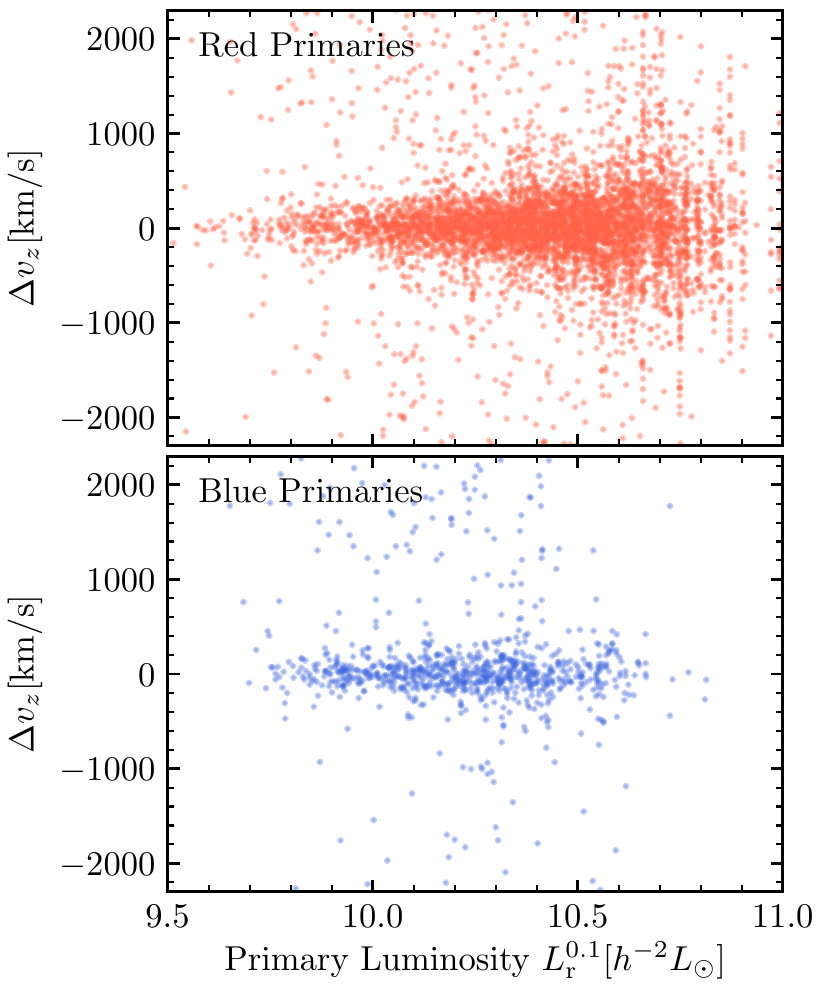}
		\caption{The velocity distribution of secondaries in SDSS DR7 as a function of the luminosity of the associated primary. We split the secondaries by the colour of the primary, red (upper panel) or blue (lower panel). It is apparent that secondaries around red primaries have a higher velocity dispersion than satellites around blue centrals of the same luminosity. Also note that red primaries also have more secondaries on average.}
		\label{fig:hist}
	\end{figure}
	
	Figure \ref{fig:hist} shows the line-of-sight velocity distribution of secondaries around primaries of a given luminosity. We show the distribution around red and blue primaries separately. From this Figure it is apparent that the velocity dispersion is a strong function of primary luminosity for red primaries and less so for blue ones \citep[see also][]{Brainerd_03, More_11, Wojtak_13}. We will analysis this more quantitatively in the coming sections.
	
	\subsection{Observables}
	\label{subsec:observables}
	
	We seek to constrain the galaxy-halo connection using a set of observables extracted from the above data set. In short, those observables are the overall number density of galaxies, the red fraction of primaries, the number of same-halo secondaries around primaries and their velocity dispersion. All quantities are measured in $10$ bins in $\log L / \Lsunh$ from $9.5$ to $11.0$.
	
	The number density of galaxies is estimated via
	\begin{equation}
		n_{\rm gal} = \frac{\sum w_{\rms, i}}{\frac{\Omega_{\rm SDSS}}{3} \left[d_{\rm com}^3 (z = 0.067) - d_{\rm com}^3 (z = 0.02)\right]},
	\end{equation}
	where $d_{\rm com}(z)$ is the comoving distance out to reshift $z$, $\Omega_{\rm SDSS} = 2.273 \mathrm{sr}$ is the effective survey area of SDSS, and the sum goes over all galaxies with a spectroscopic redshift. Finally, $w_{\rms, i}$ is a weight designed to correct spectroscopic incompleteness, as discussed in Paper I. For each galaxy, we count the number of neighbouring targets within $55''$. We then assign a weight that is the inverse of the fraction of targets with the same number of neighbours that have been assigned a spectroscopic redshift. Similarly, the red fraction of primaries is defined by
	\begin{equation}
		f_{\rm pri, r} = \frac{\sum\limits_{\rm red \ primaries}w_{\rms, i}}{\sum\limits_{\rm all \ primaries} w_{\rms, i}}.
	\end{equation}
	For the rest of the observables, we need to take into account that not all secondaries are satellites of the same halo as the primary. This is particularly important when evaluating the velocity dispersion.
	
	As discussed in detail in Paper I, we fit the $\Delta v_z$-$R_\rmp$ distribution of secondaries in each bin of primaries with a combination of an interloper $P_{\rm int} (\Delta v_z, R_\rmp)$ and same-halo satellite $P_{\rm sat} (\Delta v_z, R_\rmp)$ model, $P_{\rm tot} = P_{\rm int} + P_{\rm sat}$. Interlopers, contributing an unknown fraction $f_{\rm int}$ to all secondaries, are assumed to have a constant projected number density and a uniform distribution in line-of-sight velocities, i.e. $P_{\rm int} (\Delta v_z, R_\rmp) \propto R_\rmp$. The phase-space distribution of satellites in a halo of fixed mass is discussed in \S\ref{subsec:phase-space}. We assume that the host halo masses of satellites are drawn from a log-normal distribution with mean $\tilde{M}$ and spread $\sigma_M$, and determine the combination of $f_{\rm int}$, $\tilde{M}$ and $\sigma_M$ that maximises the likelihood
	\begin{equation}
		\log \mathcal{L} \sim \sum\limits_{\rm secondaries} w_i \log P_{\rm tot} (\Delta v_{z, i}, R_{\rmp, i})\,.
	\end{equation}
	As discussed in more detail below, there is a weight $w_i$ assigned to each primary-secondary pair. Using the resulting maximum-likelihood model, we assign each secondary a membership probability
	\begin{equation}
		p_{\rm mem} (\Delta v_z, R_\rmp) = \frac{P_{\rm sat} (\Delta v_z, R_\rmp)}{P_{\rm tot} (\Delta v_z, R_\rmp)}\,.
	\end{equation}
	We point out that the best-fitting $f_{\rm int}$, $\tilde{M}$ and $\sigma_M$ are only used to assign membership probabilities and are discarded in the subsequent analysis. 
	
	When determining the fit that maximises the likelihood, we assign a weight to each primary-secondary pair that is the product of the spectroscopic weights of the primary and the secondary;
	\begin{equation}
		w_{\mathrm{sw}} = w_{\mathrm{s, pri}} w_{\mathrm{s, scd}}\,.
	\end{equation}
	In addition to correcting for fibre collisions, this assures that each secondary receives equal weight. Thus, quantities derived from this weighting scheme are satellite-weighted. We then estimate the number of same-halo secondaries via
	\begin{equation}
		\langle N_\rms \rangle = \frac{\sum\limits_{\rm secondaries} p_{\mathrm{mem}, i} w_{\mathrm{sw}, i}}{\sum\limits_{\rm primaries} w_{\rms, i}}\,.
	\end{equation}
	Similarly, an estimate for the satellite-weighted velocity dispersion is obtained via
	\begin{equation}
		\sigma_{\rm sw}^2 = \frac{\sum\limits_{\rm secondaries} p_{\mathrm{mem}, i} w_{\mathrm{sw}, i} \Delta v_{z, i}^2}{\sum\limits_{\rm secondaries} p_{\mathrm{mem}, i} w_{\mathrm{sw}, i}}.
		\label{eq:sigma_sw}
	\end{equation}
	Finally, we wish to estimate the average velocity dispersion when weighting each primary (or host) instead of each secondary. We do so by first assigning each secondary a weight of
	\begin{equation}
		w_{\mathrm{hw}} = \frac{w_{\mathrm{s, pri}} w_{\mathrm{s, scd}}}{N_{\rm scd}},
	\end{equation}
	where $N_{\rm scd}$ is the number of secondaries hosted by the primary in question. $\sigma_{\rm hw}$ is then calculated in complete analogy to $\sigma_{\rm sw}$. Since more massive haloes typically contain more satellites, the satellite-weighted velocity dispersion typically gives relatively more weight to the more massive haloes in the luminosity bin in question. And since more massive haloes have a larger velocity dispersion, $\sigma_{\rm sw}$ will typically be larger than $\sigma_{\rm hw}$. In fact, the ratio $\sigma_{\rm sw}/\sigma_{\rm hw}$ will increase with the amount of scatter in the halo masses associated with the centrals in the luminosity bin. Thus, using both $\sigma_{\rm sw}$ and $\sigma_{\rm hw}$ as observables can constrain the scatter in halo mass at fixed luminosity \citep{More_09a}. 
	
	We measure all the above mentioned observables in $10$ logarithmic bins in luminosity going from $10^{9.5} \Lsunh$ to $10^{11} \Lsunh$. Furthermore, $\langle N_\rms \rangle$, $\log \sigma_{\rm hw}$ and $\sigma_{\rm hw}^2 / \sigma_{\rm sw}^2$ are measured for red and blue primaries separately. Ultimately, we have up to $80$ observables, $n_{\rm gal}$, $f_{\rm pri, r}$, $\langle N_{\rm s, r} \rangle$, $\langle N_{\rm s, b} \rangle$, $\log \sigma_{\rm hw, r}$, $\log \sigma_{\rm hw, b}$, $\sigma_{\rm hw, r}^2 / \sigma_{\rm sw, r}^2$ and $\sigma_{\rm hw, b}^2 / \sigma_{\rm sw, b}^2$. However, we only consider data bins for which we estimate to have at least $10$ satellites, i.e. $\sum p_{\rm mem} \geq 10$,  and for which we can create an uncertainty estimate, as described in the next section.  Note that the computation of $p_{\rm mem}$ requires an assumed radial profile, $\bar{n}_{\rm sat}(r|M)$, for the satellite galaxies. In this paper, we perform our analysis for three different choices of $\bar{n}_{\rm sat}(r|M)$, as detailed in \S\ref{subsec:phase-space} below, and the exact values of the observables differ slightly depending on which profile is adopted. As we will show in \S\ref{subsec:radial_profile}, the radial profile that best matches the observed, spatial distribution of secondaries is an NFW profile with a concentration parameter that is half that of the dark matter. Adopting that profile results in the observables listed in Table \ref{tab:data}. We emphasise, though, that in each case we compute the membership probabilities, and hence the observables, using the same radial profile as used in the subsequent analysis.
	
	\begin{table*}
		\begin{tabular}{lccccccccccc}
			\hline
			Description & \multicolumn{10}{c}{Bins} & Total\\
			\hline\hline
			$\log L_{\rm pri, min}$ & $9.50$ & $9.65$ & $9.80$ & $9.95$ & $10.10$ & $10.25$ & $10.40$ & $10.55$ & $10.70$ & $10.85$ & $9.50$ \\
			$\log L_{\rm pri, max}$ & $9.65$ & $9.80$ & $9.95$ & $10.10$ & $10.25$ & $10.40$ & $10.55$ & $10.70$ & $10.85$ & $11.00$ & $11.00$ \\
			Red Primaries & $2693$ & $3138$ & $3430$ & $3186$ & $2715$ & $1872$ & $1042$ & $408$ & $95$ & $11$ & $18590$ \\
			$\hookrightarrow$ Secondaries & $10$ & $62$ & $178$ & $410$ & $706$ & $1060$ & $1313$ & $1230$ & $734$ & $152$ & $5855$ \\
			Blue Primaries & $7161$ & $6029$ & $4888$ & $3548$ & $2317$ & $1351$ & $526$ & $126$ & $17$ & $0$ & $25963$ \\
			$\hookrightarrow$ Secondaries & $8$ & $59$ & $151$ & $179$ & $240$ & $267$ & $132$ & $63$ & $7$ & $0$ & $1106$ \\
			\hline
			$\log n_{\rm gal} [h^{3} \mathrm{Mpc}^{-3}]$ & $-2.405$ & $-2.458$ & $-2.526$ & $-2.641$ & $-2.806$ & $-3.038$ & $-3.382$ & $-3.882$ & $-4.549$ & $-5.565$ & $-1.803$ \\
			$f_{\rm pri, r}$ & $0.273$ & $0.342$ & $0.413$ & $0.474$ & $0.543$ & $0.586$ & $0.675$ & $0.776$ & $0.866$ & $1.000$ &  \\
			$\langle N_{\rm s, r} \rangle$ & -- & $0.013$ & $0.041$ & $0.096$ & $0.224$ & $0.518$ & $1.262$ & $3.413$ & $8.309$ & $16.595$ &  \\
			$\langle N_{\rm s, b} \rangle$ & -- & -- & $0.019$ & $0.038$ & $0.080$ & $0.167$ & $0.206$ & $0.438$ & -- & -- &  \\
			$\sigma_{\rm hw, r} [\kms]$ & -- & $164.1$ & $157.1$ & $163.3$ & $230.3$ & $238.2$ & $262.1$ & $402.0$ & $434.0$ & $617.3$ &  \\
			$\sigma_{\rm hw, b} [\kms]$ & -- & -- & $100.4$ & $158.8$ & $141.0$ & $181.2$ & $231.7$ & $193.2$ & -- & -- &  \\
			$\sigma_{\rm hw, r}^2 / \sigma_{\rm sw, r}^2$ & -- & -- & -- & -- & $0.851$ & $0.775$ & $0.728$ & $0.479$ & $0.589$ & $0.625$ &  \\
			$\sigma_{\rm hw, b}^2 / \sigma_{\rm sw, b}^2$ & -- & -- & -- & -- & -- & $0.692$ & $0.804$ & $0.735$ & -- & -- &  \\
			\hline
		\end{tabular}
		\caption{Overview of the data used in this analysis. We have defined $10$ bins in the luminosity of the primary $\log L_{\rm pri}$. The first two rows show the bin edges and the next four rows the number of red and blue primaries and secondaries. Finally, the last eight rows show the observables described in \S\ref{subsec:observables} that we use to constrain the model. Observables that could not be measured or for which no reliable uncertainties could be derived from mocks are omitted.}
		\label{tab:data}
	\end{table*}
	
	\section{Model Predictions}
	\label{sec:model}
	
	The ultimate goal of this study is to constrain the way in which galaxies occupy dark matter halos using the observables defined in the previous section. The functional form we use is that of a conditional luminosity function \citep[CLF][]{Yang_03}, described in detail in appendix \ref{sec:galaxy_halo_connection}. To convert a certain model of the galaxy-halo connection into a set of satellite kinematic observables, one needs an assumption about the phase-space distribution of satellites. This will be described in \S\ref{subsec:phase-space}. Ideally, one would compute all observables in mock catalogues that mimic all observational effects and directly compare those results to the SDSS. Unfortunately, this is prohibitively expensive for our current analysis. Instead, we will employ a novel approach where we use a large number of mock catalogues to find the best-fit model and an analytical approach to evaluate uncertainties. The mock catalogs are described in \S\ref{subsec:mocks} and the analytical model in \S\ref{subsec:analytical_model}.
	
	\subsection{Galaxy Phase-Space Distribution}
	\label{subsec:phase-space}
	
	Central galaxies are always assumed to reside at the dark matter halo centre and to be at rest with the bulk velocity of the halo. While it is known that central galaxies have some residual velocity dispersion with respect to the halo \citep{Behroozi_13, Guo_15a, Guo_15b, Ye_17}, the dispersion is only of the order of $15\%$ of that of the satellites. Thus, it will not strongly affect the total velocity dispersion between centrals and satellites (see Paper~I for details).
	
	Satellite galaxies follow a radial profile, $\bar{n}_{\rm sat}(r|M)$, which we assume to be described by a generalised NFW (gNFW) profile,
	\begin{equation}
		\bar{n}_{\rm sat} (r | M) \propto \left( \frac{r}{\mathcal{R} r_\rms} \right)^{-\gamma} \left( 1 + \frac{r}{\mathcal{R} r_\rms} \right)^{\gamma - 3}\,.
	\end{equation}
	Here $r_\rms$ is the scale radius of the dark matter host halo, and $\gamma$ and $\mathcal{R}$ are free parameters. For $\gamma = \mathcal{R} = 1$, satellites follow the dark matter distribution in an unbiased fashion, while their distribution becomes less centrally concentrated with increasing $\mathcal{R}$ and/or decreasing $\gamma$. In this work, we consider three choices for $(\gamma, \mathcal{R})$ to describe the SDSS data: $(1,1)$ which we call ``NFW'', $(1, 2)$ referred to as ``bNFW'' in which the satellite galaxies follow an NFW profile but with a concentration parameter that is half that of the dark matter,  and $(0, 2.5)$ which we call ``Cored''. The latter is a good fit to the radial distribution of dark matter subhaloes in the SMDPL dark-matter only simulation (see Paper~I). Together, these three profiles roughly bracket the range of profiles inferred in the literature, from the most radially concentrated \citep[e.g.,][]{Cacciato_13, Guo_15a} to the most extended \citep[e.g.,][]{Yang_05, More_09b}.
	
	We assume that satellite galaxies obey the spherically symmetric Jeans equation without velocity anisotropy. In this case, the one-dimensional velocity dispersion as a function of radius can be computed via
	\begin{align}
		\sigma^2 (r | \Vvir, c_{\rm vir}) =& \frac{c_{\rm vir} \Vvir^2}{\mathcal{R}^2 g(c_{\rm vir})} \left( \frac{r}{\mathcal{R} r_\rms} \right)^\gamma\left( 1 + \frac{r}{\mathcal{R} r_\rms} \right)^{3 - \gamma} \nonumber\\
		&\int\limits_{r / r_\rms}^{\infty} \frac{g(y)\mathrm{d}y}{(y / \mathcal{R})^{\gamma + 2} (1 + y / \mathcal{R})^{3 - \gamma}},
		\label{eq:Jeans}
	\end{align}
	\citep[e.g.,][]{vdBosch_04}. Here $\Vvir = G \Mvir / \rvir$ is the virial velocity, $c_{\rm vir} = \rvir / r_\rms$ is the virial concentration and,
	\begin{equation}
		g(y) = \ln (1 + y) - \frac{y}{1 + y}.
	\end{equation}
	
	\subsection{Mock Catalogues}
	\label{subsec:mocks}
	
	A key advantage of our study over other works using satellite kinematics is the use of detailed mock catalogues to estimate observables and their associated errors. Those mock catalogues account for all the relevant observational effects in the SDSS, and are analysed using the same pipeline as the observational data. All mock catalogues used here are based on the SMDPL dark-matter only simulation \citep{Klypin_16}. Specifically, we use the $z = 0$ ROCKSTAR \citep{Behroozi_13} halo catalogue\footnote{\url{http://yun.ucsc.edu/sims/SMDPL/hlists/index.html}}. From this catalogue, we extract all main haloes with a virial mass of $\Mvir \geq 3 \times 10^{10} \Msunh$, corresponding to $300$ times the dark matter particle mass in SMDPL. The cosmology used in SMDPL is the same as the one we assume. Furthermore, the halo mass resolution is more than sufficient to resolve galaxies above $10^{9.5} \Lsunh$ \citep[cf.,][]{Guo_15b, Sinha_18} and the volume of SMDPL is more than $10$ times bigger than the volume of SDSS we analyse. Thus, SMDPL is the ideal simulation to base our analysis on.
	
	We use \texttt{halotools} to populate the dark matter catalogue with galaxies according to the CLF recipe described in appendix \ref{sec:galaxy_halo_connection}. Specifically, for each dark matter halo, we first draw a colour for the central and afterwards a luminosity. For the satellites, we first calculate the expected occupation and then draw a number from a Poisson distribution with the same mean. Afterwards, we assign luminosities and colours based on the CLF parameters. Satellites that are brighter than their respective centrals are removed, eliminating roughly $1\%$ of all satellites. This ensures that the luminosity distribution of brightest halo galaxies (BHGs) is log-normal, in agreement with findings in SDSS \citep{Yang_08}. Note that we thereby implicitly assume that BHGs are always centrals. In \cite{Lange_18a} we have shown that this has no discernible impact on our inference. For both the centrals and satellites we consider all galaxies above a luminosity of $10^9 \Lsunh$. This threshold is lower than that of galaxies we analyse in SDSS and is chosen to have a sufficient number of targets for potential fibre collisions. Finally, the mock galaxies are assigned phase-space coordinates within their host haloes using the radial profile and Jeans equation described in \S\ref{subsec:phase-space} above.
	
	Now having a real-space galaxy catalogue, we place a virtual observer with a random position and orientation inside the simulation box. We periodically repeat the galaxy catalogue to include all galaxies within $z = 0.15$. Next, we calculate apparent magnitudes for all galaxies, taking into account average $k$ and evolution corrections, and remove all targets with $m_r > 17.6$. Subsequently, redshift space distortions and SDSS spectroscopic redshift errors are simulated by perturbing the cosmological redshift $z$ of all galaxies by
	\begin{equation}
		\Delta z = (1 + z) \times \frac{\left[v_{\rm los} + \mathcal{N}(0, \sigma_{\rm err})\right]}{c}\,.
	\end{equation}
	Here $v_{\rm los}$ is the line-of-sight velocity, and $\mathcal{N}(0, \sigma_{\rm err})$ represents a random variable drawn from a Gaussian distribution with dispersion $\sigma_{\rm err}$ centred on zero, to account for measurement errors in SDSS. Throughout we adopt $\sigma_{\rm err} = 15\kms$ \citep[][]{Guo_15b}.
	
	Subsequently, we apply the NYU VAGC \texttt{bright0} survey mask and  we simulate fibre collisions. For the latter we first construct a decollided set of galaxies from the entire sample. Within this set no two galaxies are within $55''$. The remaining galaxies are potentially collided and we randomly remove spectroscopic redshifts from $65\%$ of those. Furthermore, we remove $1\%$ of all spectroscopic redshifts in the entire sample to account for other redshift failures beside fibre collisions. In Paper I, we have shown that this fibre collision scheme captures all the salient features of spectroscopic incompleteness in the SDSS. The final step is to extract the same observables from the mock data as extracted from the real SDSS data (i.e., the observables listed in Table~\ref{tab:data}). In order to account for realisation noise, we typically construct $1000$ of such mock data sets for a given model, as detailed in \S\ref{sec:analysis} below.
	
	\subsection{Analytical Model}
	\label{subsec:analytical_model}
	
	In addition to the mock data sets described above, we will also use an analytical model to predict the various observables used here. This analytical model is based on the same assumptions regarding the galaxy-halo connection and the phase-space distribution of galaxies as in the mock catalogues. In what follows, we describe how this model is used to predict the various observables listed in Table~\ref{tab:data}.
	
	The number density of galaxies in a luminosity range $[L_1, L_2]$ can be inferred from the CLF combined with the halo mass function, $n_\rmh(M)$, via
	\begin{equation}
		n_{\rm gal}(L_1, L_2) = \int\limits_{L_1}^{L_2} \int\limits_{M_{\rm min}}^\infty \Phi_{\rm tot}(L | M) n_\rmh (M) \mathrm{d}M \mathrm{d}L.
	\end{equation}
	Here, $\Phi_{\rm tot}(L | M)$ is the combined CLF of central and satellites for a halo of virial mass $M$. The halo mass function $n_\rmh (M)$ is computed directly from the SMDPL simulation output and we assume a minimum halo mass of $3 \times 10^9 \Msunh$ to host a galaxy. See appendix \ref{sec:galaxy_halo_connection} for details regarding the CLF parametrization.
	
	The red fraction of primaries can be approximated as the red fraction of centrals,
	\begin{equation}
		f_{\rm pri, r} (L_1, L_2) \approx \frac{n_{\rm c, r} (L_1, L_2)}{n_{\rm c, r} (L_1, L_2) + n_{\rm c, b} (L_1, L_2)}.
	\end{equation}
	The number density of red (blue) centrals, $n_{\rm c, r}$ ($n_{\rm c, b}$), similarly to $n_{\rm gal}$ above, is computed by integrating the CLF of red (blue) centrals over the halo mass function. 
	
	The expected number of same-halo secondaries around primaries, $\langle N_\rms | L_1, L_2 \rangle$, can be very well described by the corresponding number of satellites around centrals that fall within the cylinder used to select the secondaries,
	\begin{align}
		\langle N_\rms | &L_1, L_2 \rangle \approx \nonumber\\ &\frac{\int\limits_{L_1}^{L_2}\int\limits_{M_{\rm min}}^\infty \langle N_\rms|M\rangle \, n_\rmh (M) \, \Phi_\rmc (L | M) \, f_{\rm ap} (L, M) \, \mathrm{d}M \mathrm{d}L}{n_\rmc (L_1, L_2)}.
	\end{align}
	Here  $\Phi_\rmc$ is the central CLF, $\langle N_\rms|M\rangle = \int_{L_{\rm th}}^\infty \Phi_\rms (L | M) \rmd L$ is the expected number of satellites above a luminosity $L_{\rm th} = 10^{9.5} \Lsunh$ for a halo of mass $M$, and $f_{\rm ap} (L, M)$ describes the fraction of satellites in a halo of mass $M$ with a central of luminosity $L$ that are expected to lie inside the cylinder defined by $R_\rms$,
	\begin{equation}
		f_{\rm ap} (L, M) = 4 \pi \int\limits_0^{\rvir} \bar{n}_{\rm sat} (r | M) \left[\zeta(r, R_\rms (L)) - \zeta(r, R_\rmc)\right] \, r^2 \, \rmd r\,,
		\label{eq:f_ap}
	\end{equation}
	where $\bar{n}_{\rm sat}(r|M)$ is the normalised radial profile of satellites in a halo of mass $M$, which obeys
	\begin{equation}
		4 \pi \int\limits_0^{\rvir} \bar{n}_{\rm sat} (r | M) \, r^2 \, \rmd r = 1\,,
	\end{equation}
	and
	\begin{equation}
		\zeta(r, R_\rms) = \begin{cases}
			1 &\quad\text{if } r \leq R_\rms \\
			1 - \sqrt{1 - R_\rms^2 / r^2} &\quad\text{otherwise.} \\ 
		\end{cases}
	\end{equation}
	Note that Eq.~(\ref{eq:f_ap}) accounts for the fact that we remove secondaries within $R_\rmc = 60 \kpch$ from their primaries, as described in \S\ref{subsec:observables}.
	
	In a similar fashion, the velocity dispersion is approximated as the expected velocity dispersion of satellites around all centrals in a given luminosity range,
	\begin{align}
		\sigma^2 &(L_1, L_2) \approx \nonumber\\
		&\frac{\int\limits_{L_1}^{L_2}\int\limits_{M_{\rm min}}^\infty w (L, M) \sigma^2_{\rm ap} (L, M) n_\rmh (M) \Phi_\rmc (L | M) \mathrm{d}M \mathrm{d}L}{\int\limits_{L_1}^{L_2}\int\limits_{M_{\rm min}}^\infty w (L, M) n_\rmh (M) \Phi_\rmc (L | M) \mathrm{d}M \mathrm{d}L}.
	\end{align}
	Here, $w (L, M)$ is a weight set to
	\begin{equation}
		w_{\rm sw} (L, M) = f_{\rm ap} (L, M) \langle N_\rms | M \rangle
	\end{equation}
	for the satellite-weighted velocity dispersion and
	\begin{equation}
		w_{\rm hw}(L, M) = 1 - \exp \left[ -f_{\rm ap}(L, M) \, \langle N_\rms | M \rangle \right]
	\end{equation}
	for the host-weighted velocity dispersion. Finally, $\sigma^2_{\rm ap} (L, M)$ is the expected average velocity dispersion of all satellites living inside a halo of mass $M$ and being inside the cylinder defined by $R_\rms(L)$
	\begin{align}
		\sigma&^2_{\rm ap} (L, M) = \nonumber\\
		&\frac{\int\limits_0^{\rvir} \bar{n}_{\rm sat} (r | M) \sigma^2 (r | M) \left[\zeta(r, R_\rms (L)) - \zeta(r, R_\rmc)\right] \, r^2 \, \rmd r}{\int\limits_0^{\rvir} \bar{n}_{\rm sat}(r|M) \left[\zeta(r, R_\rms (L)) - \zeta(r, R_\rmc)\right] \, r^2 \, \rmd r}.
	\end{align}
	\section{Analysis Procedure}
	\label{sec:analysis}
	
	The goal of this work is to constrain the galaxy-halo connection parametrized by the model\footnote{Compared to Paper I, we have added an additional free parameter to characterise the red fraction as a function of halo mass. We found this additional freedom necessary to obtain a good fit to the data. Otherwise, the model is the same as in Paper I. See appendix \ref{sec:galaxy_halo_connection} for details.} vector $\boldsymbol{\theta}$ by our observational data $\boldsymbol{D} = \{n_{\rm gal}, f_{\rm pri, r}, ...\}$. To do this we assume a likelihood of the form
	\begin{equation}
		\mathcal{L}(\boldsymbol{\theta} | \boldsymbol{D}) \propto \exp \left[ - \frac{(\boldsymbol{M}^\star (\boldsymbol{\theta}) - \boldsymbol{D})^t \boldsymbol{\Psi} (\boldsymbol{M}^\star (\boldsymbol{\theta}) - \boldsymbol{D})}{2} \right],
		\label{eq:likelihood}
	\end{equation}
	where $\boldsymbol \Psi$ is the precision matrix, i.e. the inverse of the covariance matrix, and $\boldsymbol{M}^\star (\boldsymbol{\theta}) = \langle \boldsymbol{D} (\boldsymbol{\theta}) \rangle$ the model prediction for the data vector. Both the model prediction and the covariance matrix rely on SDSS-like mock catalogues (see \S\ref{subsec:mocks} for details). However, the following simplifications are made in order to make the likelihood evaluation computationally feasible.
	
	In the \S\ref{subsec:analytical_model}, we have introduced a simple analytic model $\boldsymbol{M} (\boldsymbol{\theta})$ for the full forward-modelling prediction $\boldsymbol{M}^\star (\boldsymbol{\theta})$. In Paper I, we have shown that this simple model is able to reproduce all qualitative features of the forward-modelling prediction. At the same time, small biases between that model and results from mock catalogues of the order of $\sim 1 \sigma$ exists. Thus, we calibrate the analytic model such that it reproduces the forward-modelling prediction for a given set of parameters $\boldsymbol{\tilde \theta}$. Specifically, we introduce a bias vector $\boldsymbol{B (\boldsymbol{\tilde \theta})} = \boldsymbol{M}^\star (\boldsymbol{\tilde \theta}) - \boldsymbol{M} (\boldsymbol{\tilde \theta})$. We then approximate the forward-modelling prediction for an arbitrary model $\boldsymbol{\theta}$ as
	\begin{equation}
		\boldsymbol{M}^\star (\boldsymbol{\theta}) \approx \boldsymbol{M} (\boldsymbol{\theta}) + \boldsymbol{B (\boldsymbol{\tilde \theta})}.
		\label{eq:model}
	\end{equation}
	An estimate $\boldsymbol{\hat{B} (\boldsymbol{\tilde \theta})}$ for the bias vector is obtained from $1000$ mock catalogues created from model $\boldsymbol{\tilde \theta}$. $\boldsymbol{M}^\star$ in this case is simply the average of the $1000$ data vectors.
	
	As with the bias vector, we assume that the covariance matrix does not change significantly around any given set of parameters $\boldsymbol{\tilde \theta}$. An unbiased estimate for the covariance matrix is
	\begin{equation}
		\boldsymbol{\hat{C}} = \frac{1}{N_S - 1} \sum_i^{N_S} (\boldsymbol{D}_i - \langle \boldsymbol{D} \rangle) (\boldsymbol{D}_i - \langle \boldsymbol{D} \rangle)^t,
	\end{equation}
	where $\boldsymbol{D}_i$ is the data vector of the $i$-th out of $N_S$ mock catalogues and $\langle D \rangle$ the average of all. Finally, an unbiased estimate for the precision matrix is
	\begin{equation}
		\boldsymbol{\hat{\Psi}} = \frac{N_S - N_D - 2}{N_S - 1} \boldsymbol{\hat{C}}^{-1},
		\label{eq:precision}
	\end{equation}
	where $N_D$ is the number of data points, i.e. the dimensionality of $\boldsymbol{D}$ \citep{Taylor_13}. We note that we only use data points which can be measured in all $1000$ mock catalogues and $\sigma_{\rm hw} / \sigma_{\rm sw}$ only if $\langle N_\rms \rangle > 0.1$ on average. The latter constraint is mainly due to the distributions of $\sigma_{\rm hw} / \sigma_{\rm sw}$ in the mocks being very non-Gaussian otherwise.
	
	We use an iterative approach to derive the best-fitting model. We first create $1000$ first-generation mocks based  on the occupation model for central galaxies of \cite{More_11} and the occupation model for satellite galaxies of \cite{Cacciato_13}. These mocks, though, are a poor fit to the SDSS data, with $\chi^2 \sim 10,000$ for $56$ degrees of freedom. We now use the bias and covariance estimated from this first generation of mocks, together with the analytic model, to predict a new set of best-fitting parameters. This is done using \texttt{MultiNest} as described below. The resulting new model is predicted to have $\chi^2 \approx 90$. We then create a second generation of mocks based upon this new model. However, because the parameters of the model have changed significantly, the derived bias and covariance also change, resulting in a slightly different $\chi^2$ of $\approx 110$ for the second-generation mocks. Similarly, the analytic model combined with the updated bias and covariance now predicts a slightly different best-fitting model with $\chi^2 \approx 75$. This model is then used to create a third generation of mocks. This time, we find that the $\chi^2$ of the mocks is indeed $\sim 75$, as predicted. Similarly, the analytic model coupled with the bias and covariance of the third-generation mocks predicts a new best-fit $\chi^2$ of $\sim 70$, a negligible difference in $\chi^2$ for a $16$ parameter model. Thus, the algorithm has effectively converged to the best-fitting model and there is no need to create a new generation of mock catalogues. We then proceed to calculate the full posterior using the analytic model combined with the bias and covariance of the third-generation mocks. We refer the reader to Paper~I for a more detailed discussion of the iterative method described here, as well as a comprehensive validation of the method using mock data.
	
	For finding the maximum likelihood, as well as estimating the posterior, we make use of \texttt{MultiNest} \citep{Feroz_08, Feroz_09} and its \texttt{Python} implementation \texttt{PyMultiNest} \citep{Buchner_14}. In all cases, we use a set of $10,000$ live points and a stopping criterion of $\Delta \ln \mathcal{Z} = 10^{-4}$, where $\mathcal{Z}$ is the estimate for the global Bayesian evidence. In case of finding a best-fitting model, we run \texttt{MultiNest} in constant efficiency mode and a target efficiency of $50\%$. Although not its main purpose, \texttt{MultiNest} is an extremely robust global minimisation scheme that does not get stuck in local minima. On the other hand, when estimating the posterior, we de-activate the constant efficiency mode and set a target efficiency of $0.5\%$. To verify that our estimate for the posterior is fully converged, we have re-run \texttt{MultiNest} with $20,000$ live points and a target efficiency of $0.2\%$, getting consistent results.
	
	\section{Results}
	\label{sec:results}
	
	We have performed the analysis procedure described in the above section for three different satellite radial profiles separately. We consider satellites following dark matter in an unbiased fashion (NFW), satellites displaying an NFW profile with half the concentration of dark matter (bNFW) and a cored satellite profile (Cored). See \S\ref{subsec:phase-space} for details. In each case, we use the specific radial profile to determine likely interlopers, produce mock catalogues and fit the data. We find that the observational data we used so far cannot reliably distinguish between the different models, all giving very similar likelihoods. Consequently, we next analyse the observed radial distributions of secondary galaxies around primaries in order to distinguish better between different choices for the satellite radial profiles. Having identified the radial profile that best describes the observed radial distribution of secondaries around their host, we study in more detail the fit of the galaxy-halo connection parameters that assume this particular satellite radial profile.
	
	\subsection{Radial Profile}
	\label{subsec:radial_profile}
	
	\begin{figure*}
		\centering
		\includegraphics[width=\textwidth]{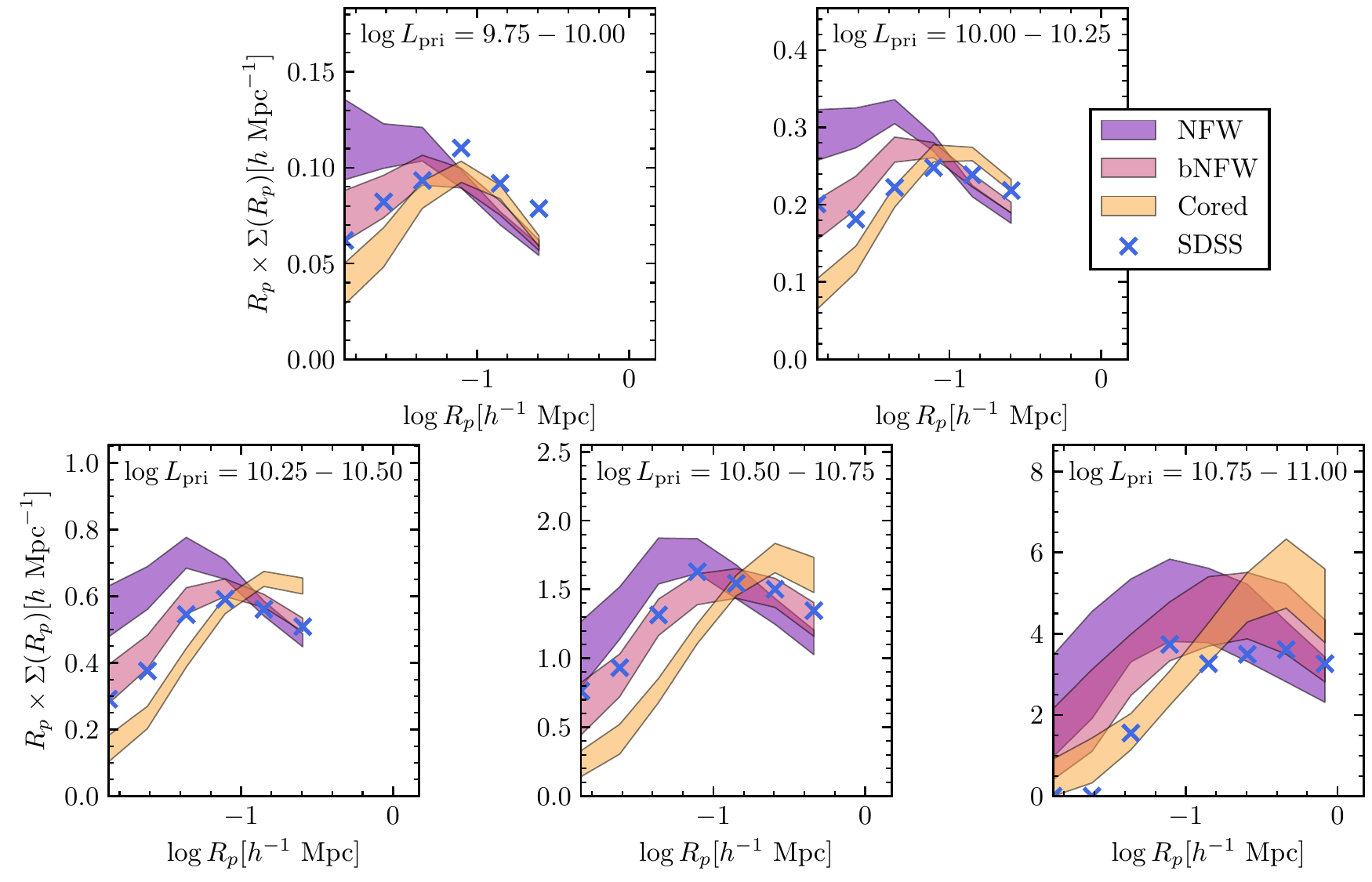}
		\caption{The projected number density of secondaries around primaries of different luminosities. Data points are from SDSS. The bands show the predictions of the best-fitting models for each assumed radial profile. For each best-fitting model, we create $1000$ mock catalogues and the bands denote the $68\%$ containment in each bin. Overall, the bNFW profile matches the SDSS data the best.}
		\label{fig:profile}
	\end{figure*}
	Figure \ref{fig:profile} shows the predictions for the projected number density of primaries around secondaries. Each panel shows the distribution for different primary luminosities without distinction by the colour of primaries. Blue crosses denote the SDSS data and bands the predictions from each of the three radial profiles. The bands present the $68\%$ scatter from $1000$ mock catalogues created with the best-fit model\footnote{Creating the mocks by drawing models from the posterior does not significantly increase the $68\%$ scatter band, indicating that cosmic variance is the dominant cause of the remaining uncertainty.}. For both the mocks and the SDSS data, we have reduced the impact of interlopers by only considering secondaries within $600 \sigma_{200}(L, C) \kms$. On the other hand, we have increased $R_\rms$ from $0.15 \sigma_{200} \Mpch$ to $0.5 \sigma_{200} \Mpch$ to increase the range that is probed. A correction for fibre collisions has been applied by weighting each primary and secondary by $w_{\mathrm{s, pri}}$ and $w_{\mathrm{sw}} = w_{\mathrm{s, pri}} w_{\mathrm{s, scd}}$, respectively.
	
	As expected, more radially extended profiles result in shallower surface number density profiles. In particular, the Cored profile predicts the lowest surface density at small radii and the largest at large radii. Overall, the bNFW profile, for which satellites follow an NFW profile with a scale radius twice that of the dark matter, fits the SDSS data the best. For example, assuming diagonal error bars, the $\chi^2$ values are $329$, $126$ and $374$ for $33$ degrees of freedom for the NFW, bNFW and Cored profiles, respectively. The largest contribution to the $\chi^2$ values comes from the lowest luminosity bin. Since the velocity dispersion is not strongly constrained for this bin, errors in the profile will not strongly influence our result. Hence, in the following we will focus on empirical results on the galaxy-halo connection derived assuming that satellite galaxies follow the bNFW radial profile. However, we emphasise that results from the other two profiles are qualitatively the same.

	\subsection{Galaxy-Halo Connection}
	
	\begin{figure*}
		\centering
		\includegraphics[width=\textwidth]{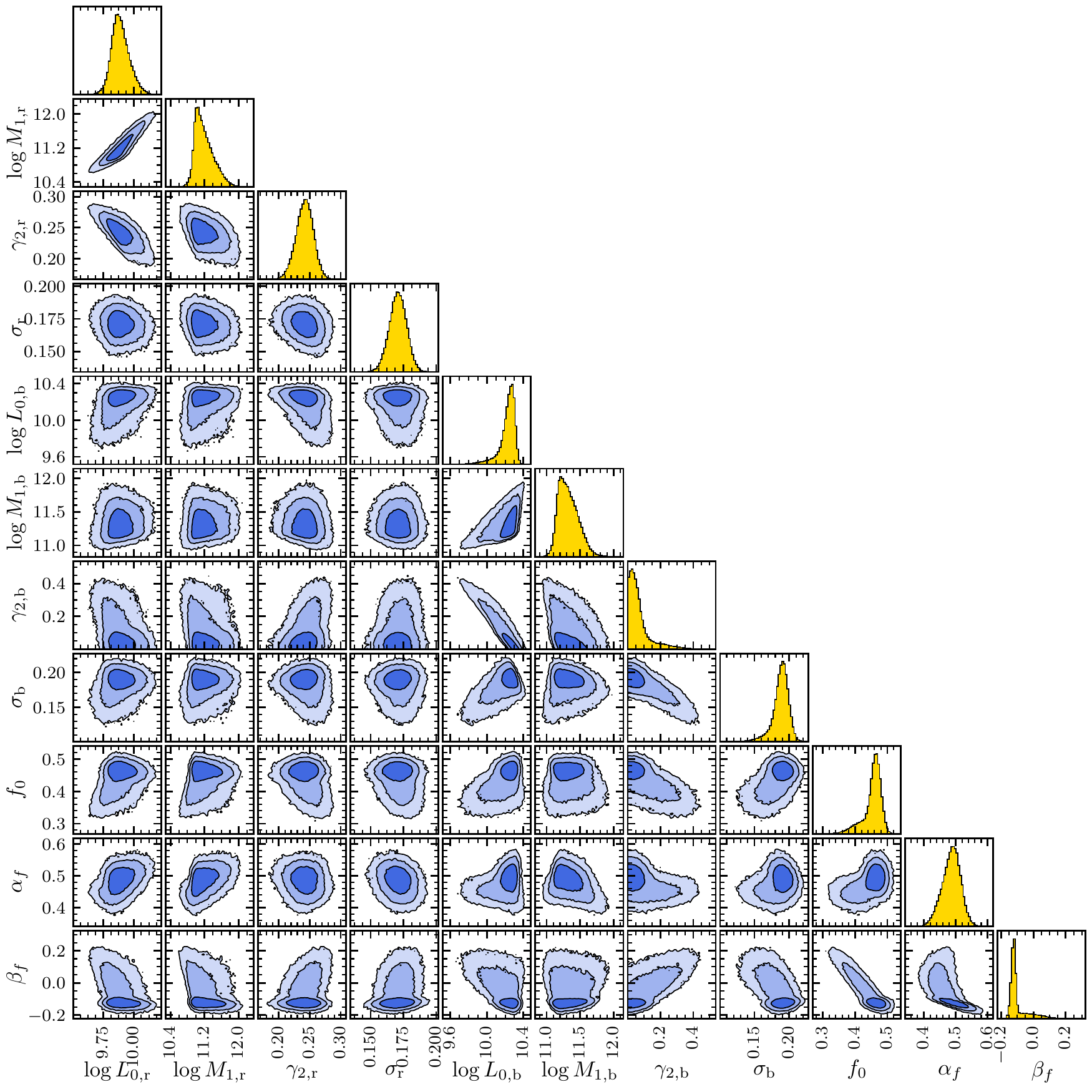}
		\caption{Marginalised posteriors for the galaxy-halo connection for the bNFW radial profile. We do not show parameters related to the satellite occupation, $\alpha_s$, $b_0$, $b_1$ and $b_2$, and the low-mass slopes of the mass-luminosity relation, $\gamma_{1, \rmr}$ and $\gamma_{1, \rmb}$, as they are poorly constrained. The diagonal shows marginalised 1D posteriors and off diagonal panels the 2D posteriors. In the latter case, lines demarcate the $68\%$, $95\%$ and $99\%$ containment of the posterior.}
		\label{fig:posterior}
	\end{figure*}
	
	\begin{table}
		\centering
		\begin{tabular}{c|c|c|c}
			\hline
			Parameter & Prior & Posterior & Best-fit \\
			\hline\hline
			$\log L_{\rm 0, r}$ & [$9.00, 10.50$] & $9.886^{+0.080}_{-0.063}$ & $9.954$ \\
			$\log M_{\rm 1, r}$ & [$10.00, 13.00$] & $11.16^{+0.27}_{-0.17}$ & $11.44$ \\
			$\gamma_{\rm 1, r}$ & [$0.00, 5.00$] & $3.27^{+1.17}_{-1.21}$ & $2.03$ \\
			$\gamma_{\rm 2, r}$ & [$0.00, 2.00$] & $0.242^{+0.013}_{-0.014}$ & $0.242$ \\
			$\sigma_{\rm r}$ & [$0.10, 0.25$] & $0.1709^{+0.0064}_{-0.0066}$ & $0.1703$ \\
			$\log L_{\rm 0, b}$ & [$9.00, 10.50$] & $10.245^{+0.051}_{-0.085}$ & $10.278$ \\
			$\log M_{\rm 1, b}$ & [$10.00, 13.00$] & $11.31^{+0.17}_{-0.13}$ & $11.51$ \\
			$\gamma_{\rm 1, b}$ & [$0.00, 5.00$] & $3.65^{+0.89}_{-0.88}$ & $2.56$ \\
			$\gamma_{\rm 2, b}$ & [$0.00, 2.00$] & $0.049^{+0.066}_{-0.033}$ & $0.036$ \\
			$\sigma_{\rm b}$ & [$0.10, 0.40$] & $0.1892^{+0.0088}_{-0.0111}$ & $0.1925$ \\
			$f_0$ & [$0.00, 1.00$] & $0.460^{+0.018}_{-0.032}$ & $0.474$ \\
			$\alpha_f$ & [$-0.50, 1.00$] & $0.488^{+0.026}_{-0.031}$ & $0.485$ \\
			$\beta_f$ & [$-0.50, 0.50$] & $-0.119^{+0.080}_{-0.017}$ & $-0.118$ \\
			$\alpha_s$ & [$-1.50, -0.90$] & $-1.13^{+0.16}_{-0.18}$ & $-0.90$ \\
			$b_0$ & [$-1.50, 0.50$] & $-0.394^{+0.038}_{-0.043}$ & $-0.355$ \\
			$b_1$ & [$0.00, 2.00$] & $0.347^{+0.080}_{-0.078}$ & $0.360$ \\
			$b_2$ & [$-0.50, 0.50$] & $0.173^{+0.027}_{-0.030}$ & $0.184$ \\
			\hline
		\end{tabular}
		\caption{Prior ranges, posterior predictions and best-fit values of all parameters describing the galaxy-halo connection. We assumed the bNFW satellite profile, i.e. that satellites follow an NFW profile with a scale radius twice as large as the dark matter scale radius.}
		\label{tab:posterior}
	\end{table}
	
	\begin{figure}
		\centering
		\includegraphics[width=\columnwidth]{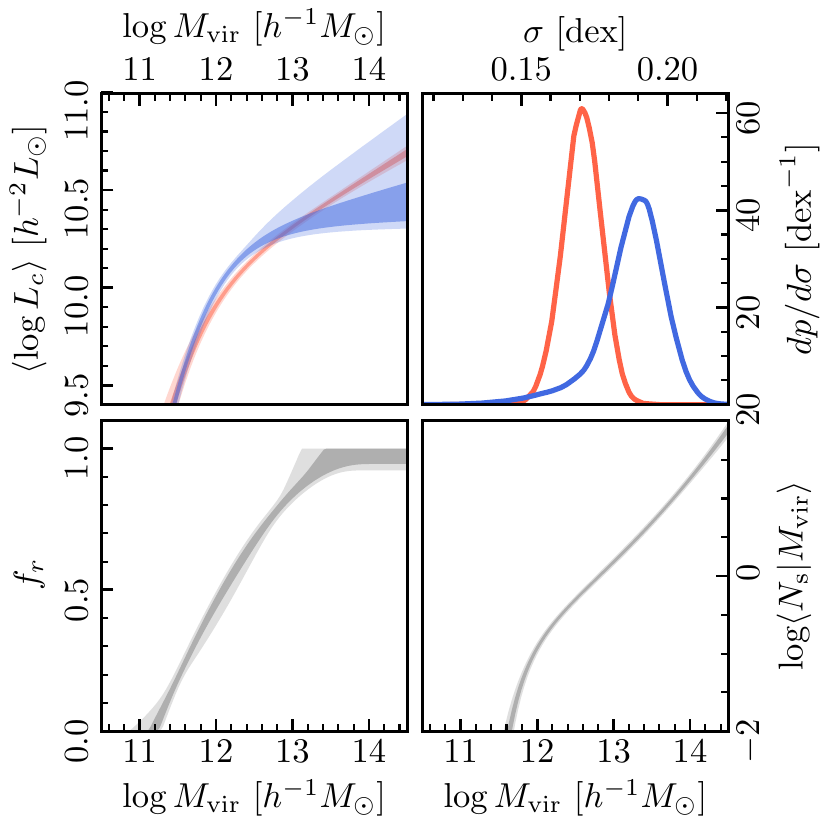}
		\caption{The marginalised posteriors on the galaxy-halo connection. We show the mass-luminosity relation for red and blue centrals (upper, left), the scatter in central luminosity at fixed halo mass for red and blue centrals (upper, right), the red fraction of centrals as a function of halo mass (lower, left) and the average number of satellites above $10^{9.5} \Lsunh$ (lower, right). Bands denote $68\%$ and $95\%$ containment ranges of the posterior.}
		\label{fig:model_post}
	\end{figure}
	
	Figure \ref{fig:posterior} shows marginalised posteriors for a subset of the parameters describing the galaxy-halo connection. Most parameters are tightly constrained with only the low-mass slopes $\gamma_{1, \rmr}$ and $\gamma_{1, \rmb}$ and the low-luminosity slope of the satellite CLF $\alpha_s$ being strongly limited by the prior. We also see tight correlations between certain parameters, for example the characteristic luminosity $L_0$ and the characteristic mass $M_1$ of the mass-luminosity relation for red and blue centrals. We list all posterior predictions on the parameters describing the galaxy-halo relation in Table \ref{tab:posterior}. We include the median and $1\sigma$ ranges as well as the best-fit value for each parameter. Figure \ref{fig:model_post} summarises the multidimensional posterior by showing predictions for key statistics of the galaxy-halo relation. Specifically, we show the predictions for the mass-luminosity relation, the scatter in luminosity at fixed halo mass, the red fraction of centrals as a function of halo mass, and the satellite occupation above $10^{9.5} \Lsunh$. Interestingly, we find mass-luminosity relations that are very similar for red and blue centrals. The same is true for the scatter in luminosity. We also find that the red fraction increases with halo mass and reaches a plateau of $\sim 95\%$ by roughly $\log M_{\rm vir} / \Msunh \sim 13$. We note that the prediction for the red fraction of lower mass haloes, $\log M_{\rm vir} / \Msunh \sim 11$, is purely an extrapolation of our model and is not constrained by data as we do not expect those haloes to host any galaxy above $10^{9.5} \ \Lsunh$. Finally, we find that the average number of satellites above $10^{9.5} \ \Lsunh$ should roughly scale linearly with the halo mass. We will compare all those results with previous studies in the next section.
	
	\subsection{Quality of the Model Fit}
	
	\begin{figure*}
		\centering
		\includegraphics[width=\textwidth]{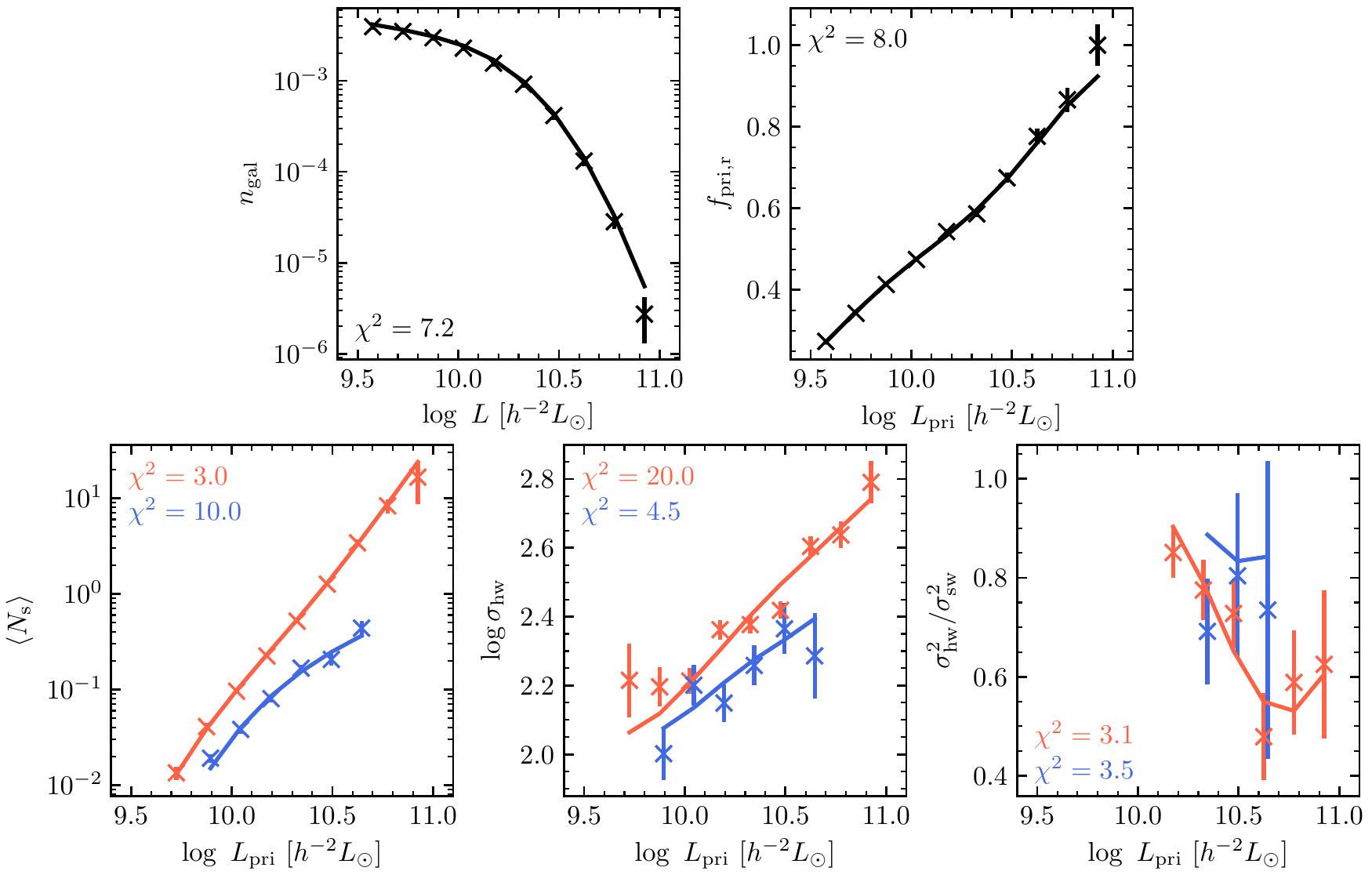}
		\caption{Observational constraints from SDSS (data points) and the best-fitting model (line). We show the number density of galaxies (top, left), the red fraction of primaries (top, right), the estimated number of satellites around primaries (bottom, left), the host-weighted velocity dispersion (bottom, middle) and the ratio of host to satellite-weighted velocity dispersion (bottom, right). All observables in the bottom row are measured separately for red and blue primaries. We also show the $\chi^2$ for each set of observations separately in each panel. Error bars represent the diagonals of the covariance matrix estimated from mock catalogues. The best-fitting model assumed the bNFW satellite profile.}
		\label{fig:constraints}
	\end{figure*}
	
	In Figure \ref{fig:constraints} we show the measurement values previously reported in Table \ref{tab:data}, together with $1\sigma$ error bars derived from mock catalogues and the best-fitting model. Qualitatively, the model excellently reproduces most trends in the observational data. However, expressed via the $\chi^2$, the best-fit model is not a very good fit to the data with $\chi^2 = 70$ for $59 - 17$ degrees of freedom. This large value for the $\chi^2$ is largely driven by the velocity dispersion of red centrals. Thus, the observational data seems to suggest a more complicated model for the galaxy-halo connection then assumed thus far. We will return to this point later in this section.
	
	\begin{figure*}
		\centering
		\includegraphics[width=\textwidth]{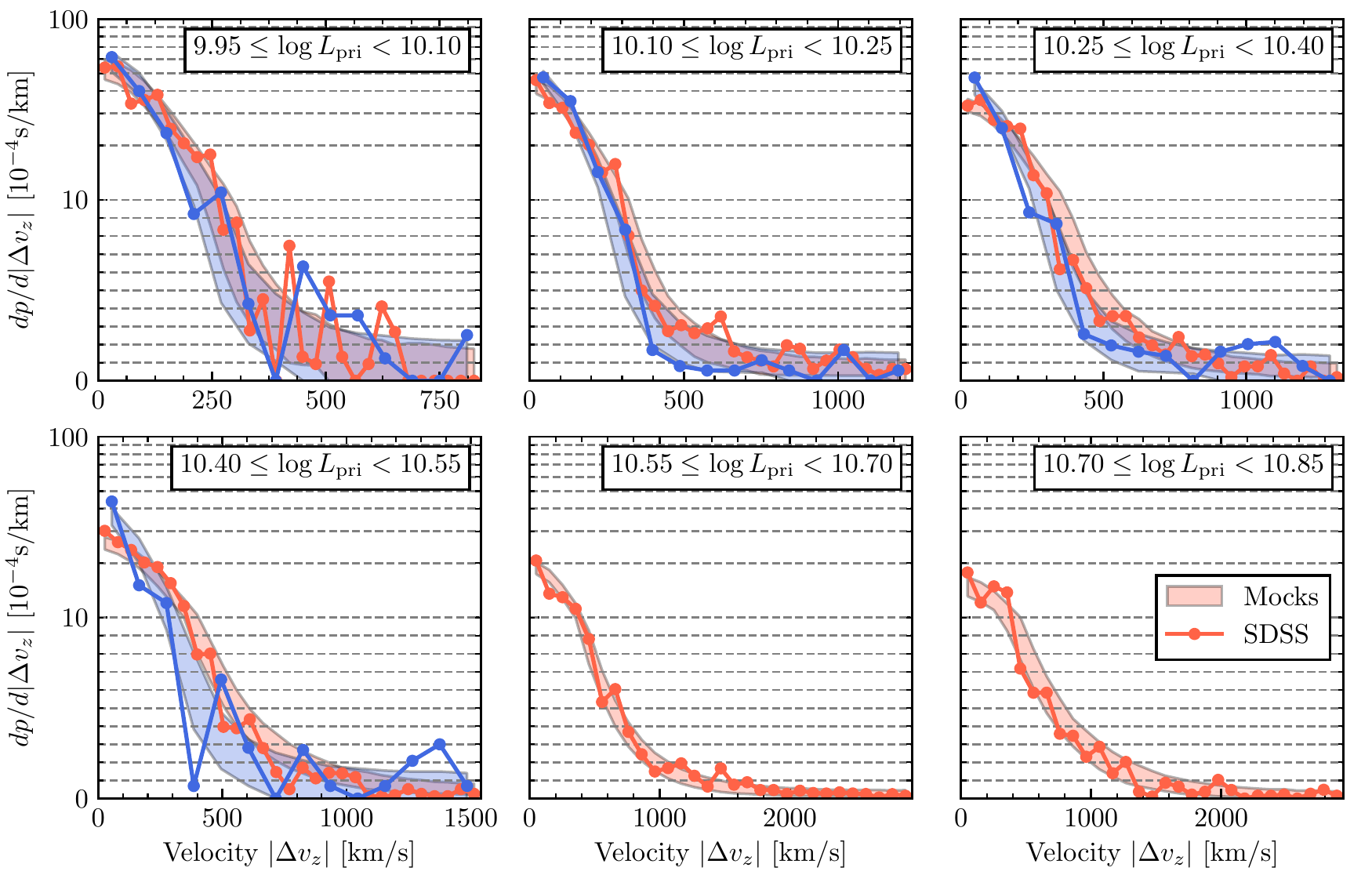}
		\caption{The velocity distribution of secondaries around primaries for SDSS (data points) and the best-fitting bNFW model (bands). Each panel represents a different primary luminosity. We show the results for red and blue primaries separately. Bands show the $68\%$ containment prediction from the best-fitting model for $1000$ mock catalogues. Note that the $y$-scale in each panel is logarithmic in the upper half and linear in the lower in order to show the full dynamic range of the data. Overall, the best-fitting model is able to reproduce all qualitative trends in the data.}
		\label{fig:vz}
	\end{figure*}
	
	In Figure \ref{fig:vz} we show the full velocity distribution of secondaries with respect to primaries. We concentrate on six different bins in primary luminosity, differentiating between secondaries around red and blue primaries. Data points come from SDSS and bands denote the $68\%$ range from $1000$ different mock catalogues with model parameters for each mock randomly drawn from the posterior. We have mitigated the effect of fibre collisions by weighting each primary-secondary pair by $w_{\mathrm{sw}} = w_{\mathrm{s, pri}} w_{\mathrm{s, scd}}$. No correction for interlopers has been applied. We note that the model has only been fitted to reproduce the second moment of the velocity distribution of satellites. On the other hand, higher order moments and the contribution of interlopers are all implicit predictions. Reassuringly, the best-fit model seems to accurately predict all of those.
	
	\subsection{Testing Model Extensions}
	
	Here, we test possible extensions to the default parametrization of the galaxy-halo connection. Specifically, we look at models with additional parameters compared to the standard parametrization. Instead of repeating the entire procedure described in \S \ref{sec:analysis}, we take the estimate of the covariance and bias from the best-fit default model and use the analytical model to evaluate the posteriors of the more complex models. This is done mainly due to computational limitations and the fact that the default model already provides an adequate description of the data. In all cases, we assume the bNFW radial profile for satellites.
	
	To judge the performance of extended models we make use of the Bayes factor $B$ \citep[see e.g.][]{Trotta_08}. In all cases, the extended model $\tilde{M}$ and the standard model $M$ can be described as nested. Nested means that the more complex model $\tilde{M}$ reduces to $M$ for $\psi$, the additional parameter of $\tilde{M}$, being $0$. In this case, one can show \citep[see e.g.][]{Trotta_08} that the Bayes factor obeys
	\begin{equation}
		B_{M \tilde{M}} = \frac{\mathcal{Z}(\boldsymbol{D} | M)}{\mathcal{Z}(\boldsymbol{D} | \tilde{M})} = \left. \frac{P(\psi | \boldsymbol{D},\tilde{M})}{P(\psi | \tilde{M})} \right|_{\psi = 0},
	\end{equation}
	where $\mathcal{Z}$ denotes the global evidence. Thus, the Bayes factor can be judged from looking at the prior, $P(\psi | \tilde{M})$, and the posterior probability, $P(\psi | \boldsymbol{D}, \tilde{M})$, of $\psi$. Values larger (smaller) than unity imply that the data $\boldsymbol{D}$ favours (disfavours) $M$ compared to $\tilde{M}$.
	
	\subsubsection{Mass-Luminosity Relation}
	
	So far, we have parametrized the mass-luminosity relation of red and blue centrals via a broken power-law relation. In $\log M$-$\log L$ space, this roughly describes a piecewise linear relation with a low and a high-mass slope. We now add to this a quadratic term in $\log M$-$\log L$ space. This can be written as
	\begin{equation}
		L_\rmc (M) = L_0 \frac{(M / M_1)^{\gamma_1}}{(1 + M / M_1)^{\gamma_1 - \gamma_2}} \left( \frac{M}{M_1} \right)^{\gamma_3 \log (M / M_1)}.
	\end{equation}
	For $\gamma_3 = 0$, this relation reduces to the previous parametrization with low-mass slope $\gamma_1$ and high-mass slope $\gamma_2$. When $\gamma_3$ is larger (smaller) than zero, though, the mass-luminosity relation has an additional upturn (downturn) at large halo masses. At the low mass end, the impact of a non-zero $\gamma_3$ is much weaker, simply because $\gamma_1 \gg \gamma_2$. Thus, we introduce $2$ new free parameters, $\gamma_{3, \rmr}$ and $\gamma_{3, \rmb}$, for red and blue centrals respectively. We use uniform priors in the range $[-0.3, +0.3]$ for both. Applying this more general model to the SDSS yields marginalised posteriors of $-0.017_{-0.021}^{+0.022}$ and $-0.095_{-0.121}^{+0.073}$ for $\gamma_{3, \rmr}$ and $\gamma_{3, \rmb}$, respectively. Thus, the data is compatible with $\gamma_{3,\rmr} = \gamma_{3,\rmb} = 0$ and the Bayes factor is $\ln B_{M \tilde{M}} = 2.8 > 0$, indicating that this additional model freedom is not favoured by the data.
	
	\subsubsection{Satellite Occupation}
	
	In principle, we might expect that at fixed halo mass the satellite occupation depends on the colour of the central galaxy. The reasoning for this is that galaxy colour might correlate with halo formation time \citep{Hearin_13c, Hearin_14}, which in turn correlates with subhalo or satellite occupation \citep{vdBosch_05b, Zentner_05, Jiang_17}. We test this by allowing the satellite CLF to vary depending on central colour, such that
	\begin{equation}
		\Phi_{\rm sat} (L | M , \mathrm{blue \ central}) = \zeta \times \Phi_{\rm sat} (L | M, \mathrm{red \ central}).
	\end{equation}
	We choose $[0.5, 2.0]$ as a (uniform) prior for this boost factor $\zeta$ and find $0.99^{+0.25}_{-0.22}$ as the posterior probability. Thus, the posterior does not exclude $\zeta = 1$ and the overall Bayer factor ($\ln B_{M \tilde{M}} = 0.9 > 0$) does not favour this model over the default parametrization. 
	
	\subsubsection{Mass-dependent scatter}
	\label{sec:massdepscatter}
	
	\begin{figure}
		\centering
		\includegraphics[width=\columnwidth]{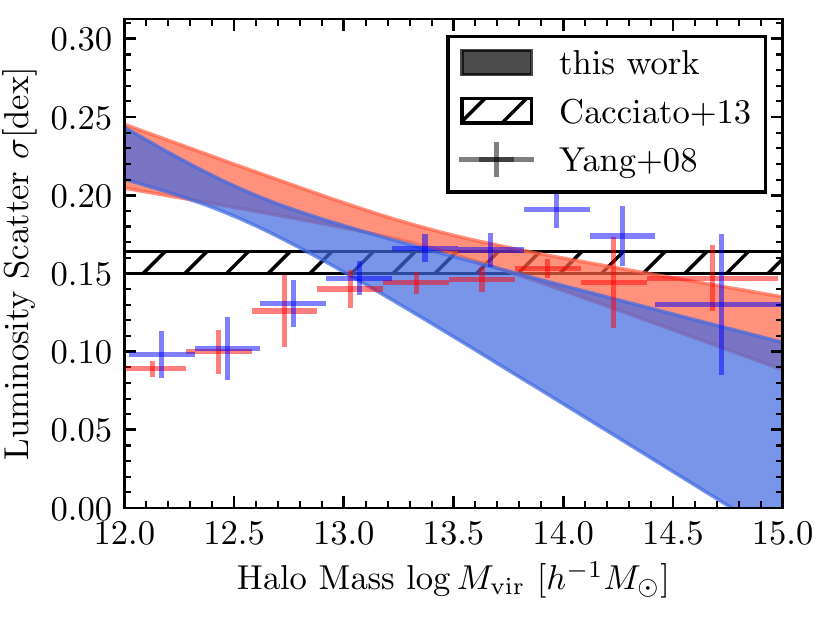}
		\caption{The mass-dependence of the luminosity scatter as inferred from the model extension discussed in \S\ref{sec:massdepscatter}. Red and blue bands denote the $68\%$ ranges from the posterior predictions for red and blue galaxies, respectively. For comparison we also show the results of \protect\cite{Yang_08}, with blue and red crosses corresponding to the results for blue and red centrals, as inferred from the SDSS galaxy group catalogue of \protect\cite{Yang_07}. Note that these results are known to underestimate the scatter at the low-mass end (see text for discussion). Finally, the mass and colour-independent inference of \protect\cite{Cacciato_13} is shown by the horizontal band, which is slightly lower than the mass-independent scatter of red or blue centrals inferred from our default model (cf. Table~\ref{tab:posterior}).}
		\label{fig:scatter_vs_mass}
	\end{figure}
	
	Finally, we allow for the scatter in luminosity to depend on halo mass. It is generally assumed that central galaxies in low-mass haloes grow via in-situ star formation whereas those in high-mass haloes do so mainly via accretion of stellar material from disrupted satellites \citep[e.g.,][]{Yang_13, Lu_15}. Thus, there is no \textit{a priori} reason that the scatter should be mass-independent \citep{Gu_16}. Indeed, hydrodynamical simulations seem to predict that the scatter increases with decreasing halo mass \citep[][]{Wechsler_18}. Note that so far we only allowed the scatter to depend on the colour of the central. We parametrize the mass dependence as follows,
	\begin{equation}
		\sigma (M) = \sigma_0 + \Delta \sigma (\log M - \log M_\sigma),
	\end{equation}
	where $\sigma_0$ and $\Delta \sigma$ are free parameters, separately for red and blue centrals, and $\log M_\sigma = 14$ ($12$) for red (blue) centrals. We choose the same uniform priors for $\sigma_{0, \rmr}$ and $\sigma_{0, \rmb}$ as for $\sigma_\rmr$ and $\sigma_\rmb$. For $\Delta \sigma_\rmr$ and $\Delta \sigma_\rmb$ we choose $[-0.1, 0.1]$. Overall, we find $\Delta \sigma_\rmr = -0.038^{+0.014}_{-0.014}$ and $\Delta \sigma_\rmb = -0.063^{+0.027}_{-0.024}$, indicating that the scatter decreases with increasing halo mass for both red and blue centrals. The mass-dependence is visualised in Figure~\ref{fig:scatter_vs_mass}. Nominally, the finding of a mass-dependent scatter is statistically significant with $\ln B_{M \tilde{M}} = - 3.0 < 0$,  strongly favouring it over the default parametrization \citep{Kass_95}. However, while we now find $\chi^2 / \mathrm{dof} = 60 / 40$, qualitatively the fit in Figure~\ref{fig:constraints} does not improve substantially. We only note that the luminosity function, specifically the value for the highest luminosity, now fits much better. Overall, we cannot exclude that other model variations besides the ones tested here might lead to a similar improvement in the fit. Thus, we only have tentative evidence that the scatter should decrease with halo mass. 
	
	Finally, we note that we have only tested generalised models for the occupation of dark matter halos with galaxies. It is possible that a generalised model for the phase-space distribution of satellites could improve the fit. For example, the radial distribution of satellites could be mass-dependent. Similarly, it is possible that there is a (mass-dependent) radial anisotropy of the satellite orbits or departures from Jeans equilibrium. Such modifications will be tested in future work.
	
	\section{Discussion}
	\label{sec:discussion}
	
	One of the main goals of this work and Paper I is to improve the analysis of satellite kinematics compared to previous efforts. Specifically, we want to investigate previously reported tensions between results from satellite kinematics and other methods \citep[see e.g.][]{Dutton_10, Leauthaud_12, Mandelbaum_16}.
	
	\subsection{Radial Profile of Satellites}
	
	Our findings shown in Figure \ref{fig:profile} favour a radial distribution of satellites that is biased with respect to dark matter. While our results are consistent with satellites following an NFW profile, their concentration parameter is likely lower than that of dark matter in the same halo by a factor of around two. At the same time, the cored profile that fits the distribution of $M_{\rm peak}$ selected subhaloes, as described in Paper I, does not provide an adequate description of the SDSS data. Thus, satellites seem to follow a steeper profile than subhaloes. 
	
	Many previous studies reported similar findings, particularly that satellite distributions follow an NFW profile \citep[see e.g.][]{vdMarel_00, vdBosch_05c, Yang_05, Budzynski_12, Guo_12a}. This is in contrast to the results of \cite{More_09b}, who argue that satellites follow a cored profile. As described in Paper I, this is likely due to \cite{More_09b} not correcting for fibre collisions in SDSS. The finding that satellites have a steeper radial profile than subhaloes but less centrally concentrated than dark matter is also in good agreement with independent studies \citep{Yang_05, Chen_08}, with hydrodynamical simulations \citep{Nagai_05, Vogelsberger_14}, and with the need for `orphan' galaxies in semi-analytical models \citep{Kitzbichler_08, Yang_12, Pujol_17} and subhalo-abundance matching \citep{Guo_14, Campbell_18, Moster_18}. Finally, we note for such comparisons, one should keep in mind that the exact radial profile of satellites will likely depend on the halo mass range and the properties of satellites due to colour and luminosity segregation \citep{Chen_08, Guo_13}.
	
	\subsection{Galaxy-Halo Occupation}
	\label{subsec:comparison}
	
	A detailed comparison with results from clustering and other methods is hampered by three complications. First, all observables like satellite kinematics or clustering will have a dependence on cosmology at fixed models for the galaxy-halo connection. Furthermore, there will be a dependence on the halo definition and the halo finder used. It is thus prudent to only compare results obtained under similar cosmological parameters and halo definitions. We compare our results to the studies of \cite{Guo_15b}, \cite{Zentner_16}, \cite{Vakili_16} and \cite{Sinha_18}. Coincidentally, all works assume cosmological parameters compatible with the recent results of \cite{Planck_14}. Even more, all studies used halo catalogues derived with the \texttt{ROCKSTAR} halo finder \citep{Behroozi_13} and spherical over-density $M_{\rm vir}$ haloes. A third complication is that all those studies use colour-independent halo occupation distribution (HOD) models. Such models only describe the number of galaxies brighter than a luminosity threshold living in a halo of mass $M$. They do not inherently describe the entire luminosity distribution of galaxies, e.g. the mass-luminosity relation, or their dependence on the colour of the central galaxy. Thus, to allow a comparison we need to downgrade our CLF results to such an HOD framework.
	
	\cite{Guo_15b} model the redshift space clustering of galaxies in SDSS using the framework of \cite{Zheng_16} to accurately and efficiently calculate correlation functions. Specifically, they model both the projected correlation function and the monopole, quadrupole and hexadecapole moments of the redshift space clustering. Furthermore, in addition to HOD parameters, they fit parameters describing the central and satellite velocity bias. \cite{Zentner_16} and \cite{Vakili_16} use \texttt{halotools} \citep{Hearin_17a} to fit the projected correlation function of galaxies in SDSS DR7. We note that the observational data used in \cite{Zentner_16} comes from \cite{Zehavi_11}, whereas \cite{Vakili_16} use data from \cite{Guo_15b}. Unlike \cite{Guo_15b}, the studies by \cite{Zentner_16} and \cite{Vakili_16} specifically allow for the presence of galaxy assembly bias, which can have a significant impact on the inferred galaxy-halo connection \citep{Zentner_14}. Finally, \cite{Sinha_18} also fit the friends-of-friends group multiplicity function in addition to the projected correlation function (both in SDSS). Their focus lies on the realistic simulation of observational effects by comparing observational results to a large number of mock catalogues. When comparing to the results of \cite{Zentner_16} and \cite{Vakili_16}, we use their models without the possibility for assembly bias. The posterior predictions from \cite{Sinha_18} discussed here are the ones using the $\Mvir$ halo definition.
	
	\begin{figure}
		\centering
		\includegraphics[width=\columnwidth]{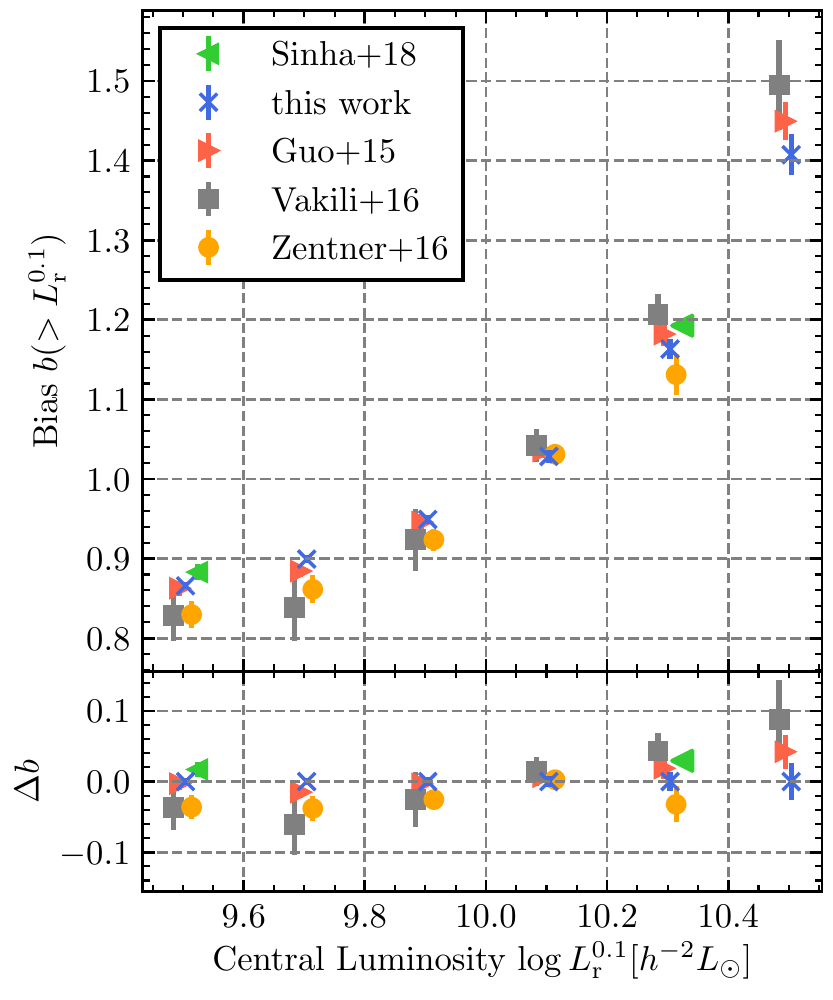}
		\caption{Prediction for the average clustering bias of central galaxies above a given luminosity (upper panel). We compare predictions from \protect\cite{Guo_15b}, \protect\cite{Vakili_16}, \protect\cite{Sinha_18}, \protect\cite{Zentner_16} and our analysis. All results include $68\%$ error bars. Predictions have been made assuming the \protect\cite{Tinker_08} halo mass function and \protect\cite{Tinker_10} model for the halo bias. We also directly compare the differences between this work and other studies at fixed luminosity (lower panel). The points are slightly offset in the horizontal direction to improve clarity. They all correspond to magnitude limits of $M_r^{0.1} - 5 \log h = -19.0, -19.5, -20.0, -20.5, -21.0$ and $-21.5$.}
		\label{fig:bias_cen}
	\end{figure}
	
	\subsubsection{Galaxy Bias}
	\label{subsec:galbias}
	
	In order to make a meaningful comparison of our results with predictions from other studies regarding the clustering of galaxies, we compute the average, linear bias of all centrals above a given luminosity threshold;
	\begin{equation}\label{galbias}
		b (> L_{\rm th}) = \frac{1}{n_\rmc (L_{\rm th})} \int\limits_0^\infty \langle N_\rmc | M, L_{\rm th} \rangle n_\rmh (M) b_\rmh(M) dM.
	\end{equation}
	Here, $n_\rmc (L_{\rm th})$ is the total number density of centrals above the luminosity threshold $L_{\rm th}$, $\langle N_\rmc | M, L_{\rm th} \rangle$ is their average number in a halo of mass $M$, and $n_\rmh(M)$ and $b_\rmh(M)$ are the halo mass and bias functions, respectively. The linear bias specifies the clustering strength on large, linear scales (in the $2$-halo regime), and allows us to  test agreement with clustering predictions despite the different parametrizations of the galaxy-halo relationship. Due to the tight relation between halo mass and bias, it is also indicative of the average mass of the haloes hosting the galaxies in question. Figure \ref{fig:bias_cen} compares our posterior prediction for $b(>L_{\rm th})$, obtained from equation~(\ref{galbias}) using the halo mass and bias function of \cite{Tinker_08} and \cite{Tinker_10}, respectively, to those of  \cite{Guo_15b} (red triangles), \cite{Zentner_16} (yellow circles), \cite{Vakili_16} (grey squares) and \cite{Sinha_18} (green triangles). All results include  $68\%$ posterior uncertainties, and slight differences in the cosmological parameters and redshifts of the halo catalogues used in the different studies are accounted for by using the corresponding halo mass and bias functions.
	
	Overall, there is excellent qualitative agreement among all studies compared here. In all works, the bias of central galaxies increases strongly with their luminosity. Given the tight correlation of halo mass with bias, this indicates a positive correlation between halo mass and galaxy luminosity. At the highest luminosities, $L_r^{0.1} = 10^{9.3} \Lsunh$ and $10^{9.5} \Lsunh$, our results predict a slightly lower central galaxy bias than those of \cite{Guo_15b}, \cite{Vakili_16} and \cite{Sinha_18}. The disagreement is generally at the level of around $2 \sigma$. Note that all clustering studies use a similar sample of galaxies drawn from SDSS. Thus, the results of those clustering studies are not statistically independent. 
	
	\subsubsection{Scatter in the luminosity-halo mass relation of central galaxies}
	\label{subsec:scatter}	
	
	Our analysis puts tight constraints on the scatter in central luminosity at fixed halo mass. Unfortunately, this is not directly constrained by results of simple HOD fitting, hampering a direct comparison of our results with the studies previously discussed. On the other hand, constraints on the colour-dependent luminosity scatter have been obtained by \cite{More_11} using satellite kinematics. They find $\sigma_\rmr = 0.21_{-0.03}^{+0.02}$ and $\sigma_{\rmb} = 0.21_{-0.06}^{+0.05}$, consistent with our results ($\sigma_r = 0.171_{-0.007}^{+0.006}$ and $\sigma_b = 0.189_{-0.011}^{+0.009}$) at the $2\sigma$-level. \cite{Yang_08} have estimated the scatter in luminosity as inferred from their group catalogues. To estimate group masses, they assume a one-to-one relationship between halo mass and the total luminosity of all group members with $M_r^{0.1} - 5 \log h < -19.5$. They find a scatter in BHG luminosity of around $\sigma \sim 0.15 \ \mathrm{dex}$ for red and blue BHGs, again in good agreement with our results. Whereas we find some evidence for an {\it increase} in scatter with decreasing halo mass (see Fig.~\ref{fig:scatter_vs_mass}), \cite{Yang_08} find their scatter to become smaller with decreasing halo mass, reaching  $\sim 0.08 \ \mathrm{dex}$ for haloes with $M \sim 2 \times 10^{12} \Msunh$. However, at such low-mass groups the central galaxy dominates the total group luminosity, and thereby the halo mass estimate, causing the inferred scatter to be artificially suppressed. \cite{Cacciato_13} have analysed the luminosity function, galaxy-galaxy lensing and clustering in SDSS in order to constrain cosmological parameters. A by-product of their analysis was a tight constraint on the scatter in luminosity at fixed halo mass equal to $\sigma_c = 0.157_{-0.007}^{+0.007}$, without making a distinction between red and blue galaxies. It is straightforward to show that this scatter for all galaxies is related to the individual scatter of red and blue galaxies via
	\begin{equation}
		\sigma_c^2 = f_r \sigma_r^2 + f_b \sigma_b^2 + 2 f_r f_b (\langle \log L_r \rangle - \langle \log L_b \rangle)^2,
	\end{equation}
	where $f_r$ and $f_b$ are the red and blue fraction of centrals at any given halo mass. As expected, the scatter is slightly increased if red and blue centrals have different average luminosities. In particular, for $\sigma_r = \sigma_b = 0.17$, $f_r = f_b = 0.5$ and $|\langle \log L_r \rangle - \langle \log L_b \rangle | = 0.1$, which is representative for the values inferred here, we get $\sigma_c = 0.184$. For other values of $f_r$ and $f_b$ (i.e. at different halo masses) the resulting scatter is slightly lower. Hence, under the assumption that the scatter is otherwise mass-independent for red and blue centrals, our results imply a slightly larger scatter than what is found by \cite{Cacciato_13}.
	\begin{figure}
		\centering
		\includegraphics[width=\columnwidth]{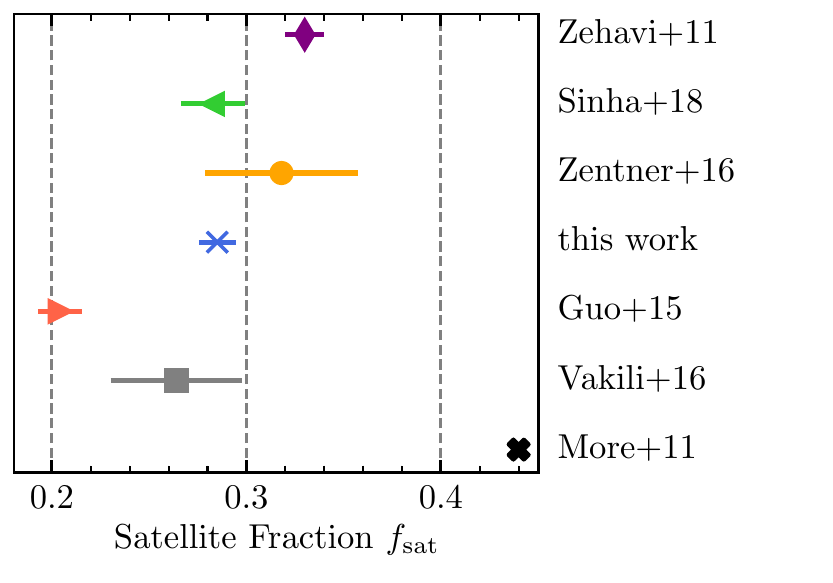}
		\caption{Predictions on the satellite fraction of galaxies brighter than $10^{9.5} \Lsunh$ for various studies. Our results, as well as those of \protect\cite{More_11}, \protect\cite{Guo_15b}, \protect\cite{Vakili_16}, \protect\cite{Zentner_16}, and \protect\cite{Sinha_18} have been derived assuming the \protect\cite{Tinker_08} halo mass function. The result for \protect\cite{Zehavi_11} is taken directly from their paper. Error bars show the $1\sigma$ uncertainty. For \protect\cite{More_11}, we show the best-fit model only.}
		\label{fig:fsat}
	\end{figure}
	
	\subsubsection{Satellite Fractions}
	\label{subsec:satfrac}	
	
	In Figure \ref{fig:fsat} we compare predictions for the satellite fraction of galaxies above $L_r^{0.1} = 10^{9.504} \Lsunh$ ($M_r^{0.1} - 5 \log h = -19$). We use the \cite{Tinker_08} mass function and account for the slightly different cosmologies and redshifts to calculate the values for our work and those of \cite{Guo_15b}, \cite{Zentner_16}, \cite{Vakili_16} and \cite{Sinha_18}. We also include the results reported in \cite{Zehavi_11}, which are, however, derived under significantly different cosmological parameters. In this comparison, we note a stronger disagreement between different predictions than in the case of the galaxy bias or the scatter discussed above. In particular, the results of \cite{Guo_15b} imply a satellite fraction of around $\sim 21 \pm 1 \%$, whereas our finding is $\sim 28 \pm 1 \%$. The results of \cite{Vakili_16} fall in between those two values, but with roughly three times larger uncertainties. \cite{Sinha_18} find a satellite fraction that is in good agreement with our findings and incompatible with those of \cite{Guo_15b}. \cite{Zentner_16} predict a similarly high fraction $\sim 32 \pm 4 \%$ but with the largest uncertainty. Finally, \cite{Zehavi_11} find $\sim 33 \pm 1 \%$, but for a different cosmology and halo definition. It is unclear at this point what the reason for the disagreement between the different studies is. We find that the mild tension between \cite{Zentner_16} and \cite{Vakili_16} is mainly caused by using different measurements for the projected correlation function. \cite{Zentner_16} use measurements from \cite{Zehavi_11}, whereas \cite{Vakili_16} use data from \cite{Guo_15b}. Given that \cite{Zentner_16} and \cite{Vakili_16} find the satellite fraction to be loosely constrained, it seems that the projected correlation function alone is compatible with a large range of satellite fractions. On the other hand, taken at face value, the redshift space clustering used by \cite{Guo_15b} seems to imply a lower satellite fraction \citep[also see][]{Guo_16} than the group multiplicity function used by \cite{Sinha_18} or the satellite kinematics used in this study. Finally, we note that the best-fit model of \cite{More_11} implies a satellite fraction of $44\%$, in clear tension with all other studies.
	
\begin{figure}
		\centering
		\includegraphics[width=\columnwidth]{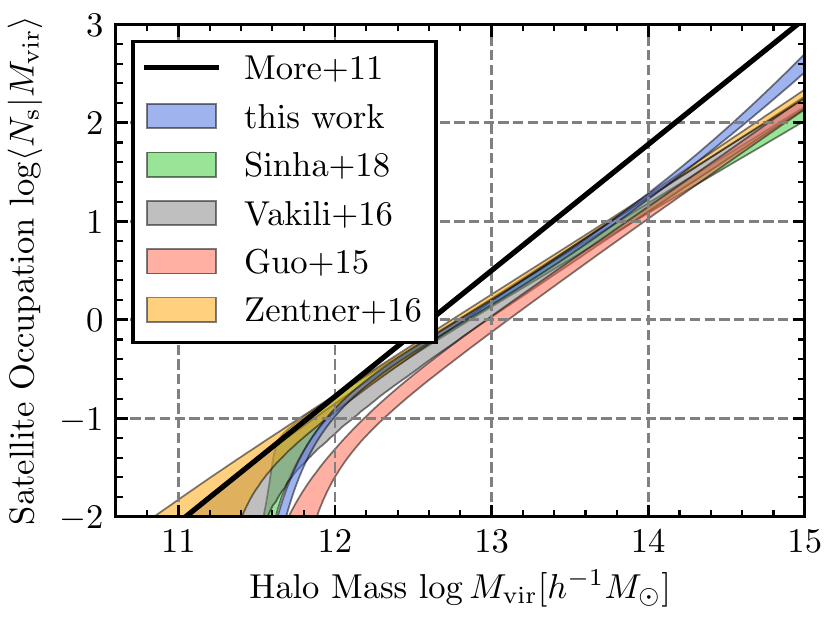}
		\caption{Posterior predictions for the satellite occupation above $10^{9.504} \Lsunh$ ($M_r^{0.1} - 5 \log h = -19$) as a function of halo mass. We compare the results of our work to those from \protect\cite{Guo_15b}, \protect\cite{Zentner_16}, \protect\cite{Vakili_16} and \protect\cite{Sinha_18}, with bands denoting the corresponding $68\%$ posterior uncertainties. We also show the best-fit model of \protect\cite{More_11}, which clearly predicts a significantly larger occupation number for
			satellites, especially in massive haloes.}
		\label{fig:nsat}
	\end{figure}

	We further explore this point in Figure \ref{fig:nsat}, where we show the satellite occupation as a function of halo mass. The threshold is $10^{9.504} \Lsunh$ ($M_r^{0.1} - 5 \log h = -19$) in all cases. Qualitatively, the results all agree very well with each other. However, at lower masses, $\log \Mvir / (\Msunh) \sim 12 - 13$, the results of \cite{Guo_15b} imply a lower satellite occupation than our findings and those of \cite{Zentner_16} and \cite{Sinha_18}. On the other hand, at the high mass end, $\log \Mvir / (\Msunh) \gtrsim 14.5$, we find a higher satellite occupation than all other studies.
	
	Altogether, we find a reasonably good, but not perfect, agreement of our results with a variety of different studies. The small discrepancies found can be explained by one or more of the following reasons. In addition to systematic uncertainties affecting satellite kinematics, as discussed in Paper~I, there are also uncertainties affecting the modelling of clustering and group multiplicity functions. A prime factor could be assembly bias, the fact that the spatial clustering of dark matter haloes depends on secondary halo properties other than halo mass \citep{Gao_05, Wechsler_06}. If the galaxy occupation correlates with any of those secondary properties, a possibility commonly ignored in HOD modelling, predictions for galaxy clustering will fail \citep[see e.g.][]{Zentner_14, Hearin_17a}. Another uncertainty is the radial profile of satellites. For example, all other studies assumed that satellites follow the dark matter radial profile in an unbiased fashion. Our results, on the other hand, imply that satellite are anti-biased with respect to dark matter. Another uncertainty is related to the second moment of the occupation statistics, which governs the small-scale clustering strength (i.e., the $1$-halo term). Although it is standard practice to assume that the number of satellite galaxies in haloes of given mass follows Poisson statistics, such that $\langle N(N-1) |M \rangle = \langle N|M \rangle^2$, this is not supported by the occupation statistics of dark matter subhaloes, which show non-negligible deviations from Poisson \citep[e.g.,][]{Boylan-Kolchin_10, Jiang_17}. Furthermore, \cite{Guo_15b}, \cite{Vakili_16} and \cite{Sinha_18} all assumed the $5$-parameter model of \cite{Zheng_07} to model observational data. However, as argued by \cite{Sinha_18}, this model might not be flexible enough to fit a variety of observations. Finally, we note that the agreement between satellite kinematics and clustering likely depends on the cosmological parameters, as it does for clustering and lensing \citep{Cacciato_09, Cacciato_13}. Thus, we might expect an agreement between all those studies only under the right cosmological parameters.
	
	\section{Conclusion and Summary}
	\label{sec:conclusion}
	
	We have obtained new constraints on the galaxy-halo relationship from the analysis of satellite kinematics in SDSS DR7. Specifically, we constrained a colour-dependent CLF model for galaxies with $0.02 \leq z \leq 0.067$ and $L_r^{0.1} \geq 10^{9.5} \Lsunh$. We made use of the updated analysis framework developed in Paper I that has been extensively tested to give unbiased results. Our results supersede the ones published in \cite{More_11} by properly accounting for observational biases, most importantly fibre collisions. Furthermore, the constraints we derive are significantly more stringent. Our main findings are as follows.
	
	\begin{itemize}
		\item As shown in Figure \ref{fig:profile}, our analysis supports a scenario in which satellite galaxies follow an NFW profile with a lower concentration than that of dark matter in the same halo. However, we rule out strongly cored profiles reported in \cite{More_11} and expected from $M_{\rm peak}$-selected subhaloes. This finding is in good agreement with hydrodynamical simulations and signals the need for orphan galaxies in subhalo abundance matching.
		\item In agreement with previous studies, we find that the average luminosity of central galaxies is positively correlated with halo mass. Additionally, the average halo mass of red centrals is higher than that of blue centrals of the same luminosity. This is primarily due to a strong increase of the red fraction of centrals with dark matter halo mass, as shown in the lower-left panel of Figure \ref{fig:model_post}, not due to different mass-luminosity relationships.
		\item We infer a scatter of $\sigma_{\rm c,r} = 0.17 \ \mathrm{dex}$ and $\sigma_{\rm c,b} = 0.19 \ \mathrm{dex}$ in luminosity at fixed halo mass for red and blue centrals, respectively. If the scatter is allowed to be mass-dependent, we infer that the scatter weakly decreases with increasing halo mass for both red and blue centrals (cf. Figure \ref{fig:scatter_vs_mass}).
		\item In \S\ref{subsec:comparison}, we compare in detail our inferences to those derived from galaxy clustering, galaxy-galaxy lensing, and group catalogues. Overall, we find that satellite kinematics gives constraints that are in good agreement with those independent studies. Most importantly, the results from our updated analysis are no longer strongly discrepant with some of these independent studies \citep[see e.g.,][]{Leauthaud_12, Mandelbaum_16}, as was the case for the satellite kinematics analysis by \cite{More_11}. However, there are significant tensions regarding the overall satellite fraction in the SDSS sample. But those tensions also exist between different studies using clustering and are not limited to satellite kinematics.
	\end{itemize}
	Although our analysis gives some of the most stringent constraints on the galaxy-halo connection to date, one should keep in mind that our analysis is only based on a small volume-limited subsample of the entire SDSS. Our framework can be easily extended to include multiple volume-limited samples, thereby increasing the statistical constraining power further. It can also easily be applied to other current or future surveys like GAMA \citep{Driver_11} or DESI \citep{DESI_16}.
	
	In the near future, we plan to apply our framework to investigate further aspects of the galaxy-halo relationship, such as the stellar-to-halo mass ratio and the correlation between galaxy size and halo mass at fixed stellar mass. In addition, satellite kinematics can also be combined with additional probes of the galaxy-halo connection to constrain cosmological parameters or to investigate galaxy assembly bias.
	
	\section*{Acknowledgements}
	
	FvdB and JUL are supported by the US National Science Foundation (NSF) through grant AST 1516962. ARZ, ASV, and KDW are funded by the Pittsburgh Particle Physics, Astrophysics, and Cosmology Center (Pitt PACC) at the University of Pittsburgh and by the NSF through grant AST 1517563. ASV was additionally funded in part by the Argonne Leadership Computing Facility, which is a DOE Office of Science User Facility supported under Contract DE-AC02-06CH11357. This research was supported  by the HPC facilities operated by, and the staff of, the Yale Center for Research Computing. FvdB received additional support from the Klaus Tschira foundation, and from the National Aeronautics and Space Administration through Grant No. 17-ATP17-0028 issued as part of the Astrophysics Theory Program.
	
	This work made use of the following software packages: \texttt{matplotlib} \citep{Hunter_07}, \texttt{SciPy}, \texttt{NumPy} \citep{vdWalt_11}, \texttt{Astropy} \citep{Astropy_13}, \texttt{Cython} \citep{Behnel_11}, \texttt{halotools} \citep{Hearin_17a}, \texttt{Corner} \citep{Foreman-Mackey_16}, \texttt{MultiNest} \citep{Feroz_08,Feroz_09, Feroz_13}, \texttt{PyMultiNest} \citep{Buchner_14}, \texttt{mangle} \citep{Hamilton_04, Swanson_08} and \texttt{pymangle}\footnote{\url{https://github.com/esheldon/pymangle}}. All the above mentioned software packages helped to greatly expedite this work.
	
	We thank Hong Guo, Manodeep Sinha and Mohammadjavad Vakili for making the posteriors of their analysis available to us. Additionaly, this work greatly benefited from useful discussions with Surhud More.
	
	The authors gratefully acknowledge the Gauss Centre for Supercomputing e.V. (www.gauss-centre.eu) and the Partnership for Advanced Supercomputing in Europe (PRACE, www.prace-ri.eu) for funding the MultiDark simulation project by providing computing time on the GCS Supercomputer SuperMUC at Leibniz Supercomputing Centre (LRZ, www.lrz.de).The Bolshoi simulations have been performed within the Bolshoi project of the University of California High-Performance AstroComputing Center (UC-HiPACC) and were run at the NASA Ames Research Center.
	
	\bibliographystyle{mnras}
	\bibliography{bibliography}
	
	\appendix
	\section{Galaxy-Halo Connection}
	\label{sec:galaxy_halo_connection}
	
	Here, we briefly describe our parametrization of the galaxy-halo connection. We refer the reader to Paper I for a more in-depth discussion.
	
	The occupation of dark matter halos with galaxies is governed by a CLF model \citep[e.g.,][]{Yang_03, vdBosch_03}. Particularly, the CLF specifies the average number of galaxies with a luminosity in the range $[L - \rmd L/2, L + \rmd L/2]$ residing in a halo of mass $M$. We split the galaxy occupation into a central and satellite component,
	\begin{equation}
		\Phi_{\rm tot}(L | M) = \Phi_\rmc (L | M) + \Phi_\rms (L | M)\,.
	\end{equation}
	Throughout, subscripts `c' and `s' refer to centrals and satellites, respectively.
	
	We further split the centrals in red (subscript `r') and blue (subscript `b'), based on the $g-r$-color, according to
	\begin{equation}
		\Phi_\rmc (L | M) = f_\rmr (M) \Phi_{\rm c, r}(L | M)+ f_\rmb (M) \Phi_{\rm c, b}(L | M).
	\end{equation}
	Here, $f_\rmr (M) = 1 - f_\rmb (M)$ is the probability that a central galaxy residing in a halo of mass $M$ is red. This probability is parametrized by
	\begin{equation}
		f_\rmr (M) =
		\begin{cases}
			f_0 + \alpha_f M_{12} + \beta_f M_{12}^2 \ \ \mathrm{for} \ \ \alpha_f + 2 \beta_f M_{12} > 0,\\
			f_0 - \alpha_f^2 / 4 \beta_f \ \ \mathrm{otherwise.}
		\end{cases}
		\label{eq:colour_cen}
	\end{equation}
	where $f_0$, $\alpha_f$ and $\beta_f$ are free parameters, and $M_{12} = \log (M / [10^{12} \Msunh])$. 
	Effectively, this equation describes a quadratic function which is equal to its global extremum wherever the first derivative would be negative. This allows models where the red fraction reaches a plateau at low or high masses. Note that we also enforce $0 \leq f_\rmr (M) \leq 1$. 
	
	We assume that every halo hosts exactly one central whose luminosity is drawn from a log-normal distribution with scatter $\sigma_\rmc$,
	\begin{equation}
		\Phi_\rmc (L | M) \, \rmd L = \frac{\log e}{\sqrt{2\pi \sigma_\rmc^2}} \exp \left[ \frac{(\log L - \log L_\rmc (M))^2}{2 \sigma_\rmc^2} \right] \frac{\rmd L}{L}\,.
	\end{equation}
	The median luminosity, which can be different for red and blue centrals, is parametrized by
	\begin{equation}
		L_\rmc (M) = L_0 \frac{(M / M_1)^{\gamma_1}}{(1 + M / M_1)^{\gamma_1 - \gamma_2}}\,.
	\end{equation}
	Additionally, the scatter in central luminosity is allowed to be different for red and blue centrals. Altogether, $L_{0, \rmr}$, $M_{1, \rmr}$, $\gamma_{1, \rmr}$, $\gamma_{2, \rmr}$, $\sigma_\rmr$, $L_{0, \rmb}$, $M_{1, \rmb}$, $\gamma_{1, \rmb}$, $\gamma_{2, \rmb}$, $\sigma_\rmb$, $f_0$ and $\alpha_f$ are the free parameters governing the occupation of dark matter haloes with central galaxies.
	
	The occupation of satellite galaxies is assumed to follow a Poisson distribution with expectation value
	\begin{equation}
		\langle N_\rms|M \rangle = \int\limits_{L_{\rm th}}^\infty \Phi_\rms (L | M) \, \rmd L\,,
	\end{equation}
	where $L_{\rm th}$ is a luminosity threshold. Following \cite{Yang_08}, the CLF of satellite galaxies follows a modified Schechter function,
	\begin{equation}
		\Phi_\rms (L | M) = \phi_\rms^* (M) \left( \frac{L}{L_\rms^*(M)} \right)^{\alpha_\rms} \exp \left[ - \left( \frac{L}{L_\rms^* (M)} \right)^2 \right] \frac{\rmd L}{L}\,.
	\end{equation}
	Here $\alpha_\rms$ is a free parameter that characterizes the low-luminosity slope of the satellite CLF,
	$\phi_\rms^*(M)$ is a mass-dependent normalization, and $L_\rms^* (M)$ is the (mass-dependent) characteristic luminosity. \cite{Yang_08} found that the latter is closely related to the average luminosity of centrals according to, $L_\rms^* (M) \sim 0.562 L_\rmc (M)$. Since red centrals are more prominent in our sample than blue centrals, we therefore adopt
	\begin{equation}
		L_\rms^*(M) = 0.562 L_{\rmc, \rmr} (M).
	\end{equation}
	We follow \cite{Cacciato_09} and characterize the mass-dependent normalization of the satellite CLF using three free parameters, $b_0$, $b_1$, and $b_2$, according to
	\begin{equation}
		\log \left[ \phi_\rms^* (M) \right] = b_0 + b_1 \, M_{12} + b_2 \, M_{12}^2.
	\end{equation}
	Finally, satellites are also randomly assigned colours based on the parametrization in eq. (\ref{eq:colour_cen}) with $f_{0, \rm sat} = 0.44$, $\alpha_{f, \rm sat} = 0.14$ and $\beta_{f, \rm sat} = 0$ \citep[compare][]{Yang_08}. This colour assignment only matters in the rare cases that satellites are identified as primary candidates. Thus, we have $\alpha_\rms$, $b_0$, $b_1$ and $b_2$ as free parameters for the satellite CLF. 
	
	We note that, in principle, this CLF model allows satellites to be brighter than centrals. We explicitly remove satellites that are brighter than their respective centrals from our mock galaxy catalogues. In principle, doing so also lowers $\langle N_\rms|M \rangle$. However, this effect is very small and we neglect it in our analytical model described in \S\ref{subsec:analytical_model}.
	
	\label{lastpage}
\end{document}